\documentclass[a4paper,11pt]{article}
\pdfoutput=1 

\usepackage{jheppub2} 
\usepackage{bbm,bm,graphicx,mathtools,color,hyperref,slashed}
\usepackage{amssymb,amsmath}

\usepackage[all]{xy}
 \usepackage[caption=false]{subfig}

\def\be{\begin{equation}}
\def\ee{\end{equation}}
\def\bea{\begin{eqnarray}}
\def\eea{\end{eqnarray}}
\def\bequ{\begin{equation}}
\def\eequ{\end{equation}}

\def\Tr{\mbox{Tr}}
\def\del{\partial}

\newcommand{\beq}{\begin{eqnarray}}
\newcommand{\eeq}{\end{eqnarray}}
\newcommand{\eq}{equation}
\newcommand{\eqa}{eqnarray}

\newcommand{\diff}{\mathrm{d}}
\newcommand{\p}{\partial}
\newcommand{\ve}{\varepsilon}
\newcommand{\im}{\mathrm{i}}
\newcommand{\Diff}{{\mathcal{D}}}
\newcommand{\rme}{\mathrm{e}}
\newcommand{\mlat}{m_{\mathrm{lat}}}

\usepackage{mathtools}

\DeclarePairedDelimiter\ket{\lvert}{\rangle}
\DeclarePairedDelimiterX\braket[2]{\langle}{\rangle}{#1 \delimsize\vert #2}

\title{DMRG study of the higher-charge Schwinger model and its 't Hooft anomaly}

\author[1,2]{Masazumi Honda,}
\affiliation[1]{
Yukawa Institute for Theoretical Physics, Kyoto University, Sakyo-ku, Kyoto 606-8502, Japan}
\emailAdd{masazumi.honda(at)yukawa.kyoto-u.ac.jp}

\author[2,3,4]{Etsuko Itou,}
\affiliation[2]{Interdisciplinary Theoretical and Mathematical Sciences Program (iTHEMS), RIKEN, Wako 351-0198, Japan}
\affiliation[3]{Department of Physics, and Research and 
Education Center for Natural Sciences, Keio University, 4-1-1 Hiyoshi, Yokohama, Kanagawa 223-8521, Japan}
\affiliation[4]{Research Center for Nuclear Physics (RCNP), Osaka University, Osaka 567-0047, Japan}
\emailAdd{itou(at)yukawa.kyoto-u.ac.jp}

\author[1]{Yuya Tanizaki}
\emailAdd{yuya.tanizaki(at)yukawa.kyoto-u.ac.jp}

\abstract{
The charge-$q$ Schwinger model is the $(1+1)$-dimensional quantum electrodynamics (QED) with a charge-$q$ Dirac fermion. 
It has the $\mathbb{Z}_q$ $1$-form symmetry and also enjoys the $\mathbb{Z}_q$ chiral symmetry in the chiral limit, and there is a mixed 't Hooft anomaly between those symmetries. 
We numerically study the charge-$q$ Schwinger model in the lattice Hamiltonian formulation using the density-matrix renormalization group (DMRG). 
When applying DMRG, we map the Schwinger model to a spin chain with nonlocal interaction via Jordan-Wigner transformation, and we take the open boundary condition instead of the periodic one to make the Hilbert space finite-dimensional. 
When computing the energy density or chiral condensate, we find that using local operators significantly reduces the boundary effect compared with the computation of corresponding extensive quantities divided by the volume. 
To discuss the consequence of the 't~Hooft anomaly, we carefully treat the renormalization of the chiral condensates, and then we confirm that Wilson loops generate the discrete chiral transformations in the continuum limit. 
}

\preprint{YITP-22-100, RIKEN-iTHEMS-Report-22}

\begin{document}
\maketitle

\section{Introduction}

The charge-$q$ Schwinger model~\cite{Anber:2018jdf, Anber:2018xek, Armoni:2018bga, Misumi:2019dwq} 
is the $(1+1)$-dimensional $U(1)$ gauge theory coupled to a Dirac fermion with the gauge charge $q$. 
Here, $q$ must be an integer to satisfy the large $U(1)$ gauge invariance 
and this model has the $\mathbb{Z}_q$ $1$-form symmetry
that is a remnant of $U(1)$ 1-form symmetry in the pure Maxwell theory.
Although the local dynamics of this model 
is identical to that of the usual Schwinger model~\cite{Schwinger:1962tp} with $q=1$, 
the presence of the $1$-form symmetry gives the crucial difference in the global aspects between these models~\cite{Pantev:2005rh, Pantev:2005wj, Pantev:2005zs, Hellerman:2006zs, Hellerman:2010fv}. 
Especially when we take the chiral limit, 
the charge-$q$ model has the $\mathbb{Z}_q$ discrete chiral symmetry, which we denote $(\mathbb{Z}_q)_\chi$, 
while the usual one does not have it at all due to the Adler-Bell-Jackiw (ABJ) anomaly \cite{Adler:1969gk, Bell:1969ts}. 

It is known that 
there is a mixed 't~Hooft anomaly between the $\mathbb{Z}_q^{[1]}$ and $(\mathbb{Z}_q)_\chi$ symmetries 
so that the anomaly matching condition predicts $q$ distinct vacua as a consequence of the spontaneous breaking of the discrete chiral symmetry $(\mathbb{Z}_q)_\chi$. 
It enables us to understand many dynamical consequences of the Schwinger model studied in '70s--'90s~\cite{Lowenstein:1971fc, Coleman:1975pw, Coleman:1976uz, Manton:1985jm, Faddeev:1986pc, Jayewardena:1988td, Hetrick:1988yg, Iso:1988zi, Smilga:1992hx, Adam:1993fc, Adam:1997wt} solely in the kinematical way thanks to the language of generalized symmetry \cite{Gaiotto:2014kfa}. 
The development of generalized symmetries has expanded the applicability of the 't~Hooft anomaly matching condition, 
and we nowadays find that 
many new anomalies give the constraints on strongly-coupled field theories 
including the $4$d gauge theories with massless fermions~\cite{Gaiotto:2017yup, Tanizaki:2017bam, Komargodski:2017dmc, Komargodski:2017smk, Shimizu:2017asf, Gaiotto:2017tne}, such as massless quantum chromodynamics (QCD). 
Furthermore, a recent study shows that $4$d QCD-like theories can be reduced to the $2$d QFTs keeping their anomaly constraints~\cite{Tanizaki:2022ngt, Tanizaki:2022plm} 
(see also Refs.~\cite{Yamazaki:2017ulc, Tanizaki:2017qhf, Tanizaki:2017mtm, Yamazaki:2017dra, Cox:2021vsa}), 
and these anomalies turn out to be almost identical to the 't Hooft anomaly of the charge-$q$ massless Schwinger model. 
These developments motivate us to reconsider the charge-$q$ Schwinger model for a better understanding of the microscopic structure of anomaly matching.

We numerically analyze the charge-$q$ Schwinger model on a lattice to study the above phenomena in this paper. 
For this purpose, we should investigate parameter regime with non-small vacuum angle $\theta$ 
which causes the infamous sign problem in the standard Monte Carlo lattice simulation.\footnote{
See \cite{Fukaya:2003ph} for earlier work on the Schwinger model with theta term by a Monte Carlo approach.}
There are several options to circumvent this issue, such as the lattice Hamiltonian formulation~\cite{Banks:1975gq, Carroll:1975gb, Hamer:1997dx, Banuls:2013jaa, Rico:2013qya, Banuls:2016lkq, Buyens:2016hhu,  Funcke:2019zna,Chakraborty:2020uhf,Honda:2021aum,Honda:2021ovk}, tensor renormalization group~\cite{Shimizu:2014uva, Shimizu:2014fsa}, or the use of dual variables with the Villain-type Euclidean lattice~\cite{Gattringer:2018dlw,Sulejmanpasic:2019ytl}. 
Here we adopt the Kogut-Susskind Hamiltonian formalism \cite{Kogut:1974ag} with the staggered fermion in this paper. 
We take the open boundary condition instead of the periodic one to make the physical Hilbert space finite-dimensional, 
and map the system to a spin chain with a nonlocal interaction by the Jordan-Wigner transformation. 
We apply the density-matrix renormalization group (DMRG) to study the low-energy properties of this system, using the ITensor Library~\cite{itensor}. 

As a price of the open boundary condition, physical observables receive effects of boundaries.
In particular,
the open boundary condition violates the periodicity of the vacuum angle, $\theta\sim \theta+2\pi q$,
while it holds for closed spacetime both on the continuum and lattice. 
As we need to study physics at large values of $\theta$, 
this explicit violation of the $\theta$ periodicity is an undesirable feature. 
We note that the $1$-flavor Schwinger model is gapped even in the chiral limit so that the effect of the boundary should become exponentially small when we probe local operators. 
Therefore, we compute energy density, scalar condensate, and pseudo-scalar condensate using local operators after their UV renormalization. We carefully analyze the chiral condensate in the continuum limit and show that it draws a circle around the origin as the $\theta$ parameter is changed gradually from $0$ to $2\pi q$.

To confirm the consequence of the mixed 't~Hooft anomaly, we compute the local observables under the presence of Wilson loops, or equivalently external electric charges. 
In the chiral limit, the energy densities inside and outside of the probe charge are the same, which shows that the Wilson loop becomes topological in the long-range limit. The phase of the chiral condensates rotates by $2\pi/q$ across the external charge, so it implies that the Wilson loop can be regarded as a generator for the discrete axial transformation for infrared observers. It is nothing but the consequence of the 't~Hooft anomaly for the charge-$q$ Schwinger model. 

This paper is organized as follows.
In sec.~\ref{sec:Schwinger_review} we review the charge-$q$ Schwinger model.
In sec.~\ref{sec:lattice}, we present the lattice formulation of the Schwinger model
and explain how to compute observables.
In sec.~\ref{sec:results}, we present our simulation results.
Sec.~\ref{sec:summary} is devoted to summary and discussion.

\section{Review on the charge-\texorpdfstring{$q$}{q} Schwinger model}
\label{sec:Schwinger_review}

In this section, we give a brief review on the charge-$q$ Schwinger model~\cite{Anber:2018jdf, Anber:2018xek, Armoni:2018bga, Misumi:2019dwq}. 
Although the local property is exactly identical to the usual Schwinger model, its global aspect turns out to be more fruitful than the usual one. 
Especially, the chiral limit of the charge-$q$ Schwinger model enjoys the discrete chiral symmetry $(\mathbb{Z}_q )_\chi$ and the ground states are $q$-fold degenerate due to its spontaneous breaking. 

\subsection{Charge-\texorpdfstring{$q$}{q} Schwinger model and its \texorpdfstring{$\mathbb{Z}_q^{[1]}$}{Zq 1-form} symmetry}
The Euclidean action of the charge-$q$ Schwinger model is given by\footnote{
In this section we use the differential form notation since it is convenient to discuss 't Hooft anomalies and topological aspects. The Lagrangian density for the gauge field in the component notation is
$\frac{1}{4g^2}F_{\mu\nu}F^{\mu\nu} -\frac{\im \theta}{4\pi}\epsilon_{\mu\nu} F^{\mu\nu}$.
} 
\begin{equation}
S
=\int_{M_2}\left(\frac{1}{2g^2}F\wedge \star F-\frac{\im\:\! \theta}{2\pi}F\right)+\int_{M_2}\overline{\psi}\left( \gamma^\mu_{\mathrm{E}}(\partial_\mu+\im\:\! q A_\mu)+m\right)\psi\, \diff^2 x, \label{eq:Euclid-action}
\end{equation}
where $A$ is the $U(1)$ gauge field, $F=\diff A$ is the gauge-field strength, $\psi$ is the fermion field, and $\gamma^\mu_{\mathrm{E}}$ is the Euclidean gamma matrix satisfying $\{\gamma^\mu_{\mathrm{E}},\gamma^{\nu}_{\mathrm{E}}\}=2\delta^{\mu\nu}$. 
The $U(1)$ charge of the fermion is given by an integer $q\in \mathbb{Z}$. 
This quantization comes from the fact that the (large) $U(1)$ gauge transformation on closed spacetimes is given by
\begin{equation}
    A\mapsto A+\diff \lambda,\quad \psi\mapsto \rme^{-\im\:\! q\lambda}\psi, 
\end{equation}
with $\rme^{\im \lambda}:M_2\to U(1)$, and this is unambiguous only if $q\in \mathbb{Z}$.
Accordingly, the $U(1)$ gauge field $A$ satisfies the Dirac quantization condition, $\int_{M_2} \diff A \in 2\pi \mathbb{Z}$, on any oriented closed $2$-manifolds, 
and thus the $\theta$ angle becomes the $2\pi$ periodic parameter on closed spacetime. 

When $q>1$, the theory enjoys the $\mathbb{Z}_q$ $1$-form symmetry, which we denote as $\mathbb{Z}_q^{[1]}$. 
This is a remnant of the $U(1)$ 1-form symmetry in the pure Maxwell theory
partially broken by inclusion of the charge-$q$ matter to $\mathbb{Z}_q$.
The presence of the $\mathbb{Z}_q$ $1$-form symmetry implies that
the theory is completely decomposed into $q$ distinct sectors 
called universe~\cite{Pantev:2005rh, Pantev:2005wj, Pantev:2005zs, Hellerman:2006zs, Hellerman:2010fv}
and each sector has a direct relation to the usual Schwinger model
as we will review below.

A convenient way to observe this fact is to turn on a background gauge field 
for the $\mathbb{Z}_q$ 1-form symmetry that is given by a $\mathbb{Z}_q$-valued 2-form gauge field.
For this purpose, we realize it as a pair of the $U(1)$ $2$-form gauge field $\mathcal{B}$ and the $U(1)$ $1$-form gauge field $\mathcal{C}$, with the constraint 
\begin{equation}
    q\:\! \mathcal{B}=\diff \mathcal{C}. 
\end{equation}
Due to this constraint, the holonomy of $\mathcal{B}$ is quantized in $\mathbb{Z}_q\subset U(1)$, and it becomes the $\mathbb{Z}_q$ gauge field. 
The $2$-form gauge field has the $U(1)$ $1$-form gauge transformation, 
\begin{equation}
    \mathcal{B}\mapsto \mathcal{B}+\diff \Lambda, \quad 
    \mathcal{C}\mapsto \mathcal{C}+q \Lambda, 
\end{equation}
where the gauge parameter $\Lambda$ itself is the $U(1)$ $1$-form gauge field. 
Under this background gauge transformation, we impose that
the dynamical gauge field $A$ is transformed as
\begin{equation}
    A\mapsto A-\Lambda .
\end{equation}
An action invariant under the above transformation with a minimal coupling is
\begin{align}
    S_{\mathrm{gauged}} [\mathcal{B}]
    &=\int_{M_2}\left(\frac{1}{2g^2}(F+\mathcal{B})\wedge \star (F+\mathcal{B})-\frac{\im\:\! \theta}{2\pi}(F+\mathcal{B})\right)\nonumber\\
    &\quad +\int_{M_2}\overline{\psi}\left( \gamma^\mu_{\mathrm{E}}(\partial_\mu+\im (q A_\mu+\mathcal{C}_\mu))+m\right)\psi\, \diff^2 x .
\end{align}
By performing the path integral for dynamical fields, 
we obtain the partition function with the background gauge field $\mathcal{B}$ as
\begin{equation}
Z_q [g,\theta ; \mathcal{B}]
:=\int \Diff A \Diff \overline{\psi}\Diff \psi 
\exp\left(-S_{\mathrm{gauged}}[\mathcal{B} ] \right) .
\end{equation}

While we can go back to the original partition function by simply taking $\mathcal{B}=0$, 
as we will see below,
we can relate the charge-$q$ Schwinger model to the usual $q=1$ case
if we do ``taking $\mathcal{B}=0$" in a bit sophisticated way.
This is done by performing the path integral over $\mathcal{B}$ with an appropriate weight and summing over the weight.
In this process, we consider the following topological action
\begin{equation}
    \im k \int_{M_2}  \mathcal{B}. 
\end{equation}
Under the $U(1)$ $1$-form gauge transformation, this is gauge invariant modulo $2\pi \im k\mathbb{Z}$, and thus it is well-defined if $k$ is an integer. 
This quantity is quantized in $\frac{2\pi \im k}{q}\mathbb{Z}$, so $k$ is identified with $k+q$, i.e. $k\in \mathbb{Z}_q$. 
Let us perform the path integral over $\mathcal{B}$ including its discrete topological term:
\begin{align}
\int \Diff \mathcal{B} Z_q[\theta ;\mathcal{B}]\rme^{\im k\int \mathcal{B}} 
    &=\int \Diff \mathcal{B} \Diff \mathcal{C}\int \Diff A \Diff \overline{\psi}\Diff \psi \exp\left(-S_{\mathrm{gauged}}[\mathcal{B}]+\im k \int \mathcal{B}\right)\nonumber\\
    &=\int \Diff \mathcal{C} \Diff \overline{\psi}\Diff \psi \exp\left(-S'\right), 
    \label{eq:universe_Z}
\end{align}
where 
\begin{equation}
    S'=\int_{M_2}\left(\frac{1}{2(qg)^2}\diff \mathcal{C}\wedge \star \diff \mathcal{C}+\im\frac{\theta+2\pi k}{2\pi q}\diff \mathcal{C}\right)+\int_{M_2}\overline{\psi}\left( \gamma^\mu_{\mathrm{E}}(\partial_\mu+\im \mathcal{C}_\mu)+m\right)\psi\, \diff^2 x. 
\end{equation}
This is nothing but the charge-$1$ Schwinger model, with the gauge coupling $g'=qg$ and the vacuum angle $\theta'=(\theta+2\pi k)/q$. 
Therefore we find
\begin{equation}
\int \Diff \mathcal{B} 
  Z_q[g,\theta ;\mathcal{B}]\rme^{\im k\int \mathcal{B}}
= Z_1 \left[ qg, \frac{\theta +2\pi k}{q}   ;\mathcal{B}=0 \right] .
\end{equation}
We can undo the $\mathbb{Z}_q^{[1]}$ gauging by summing over the discrete labels $k=0,1,\ldots, q-1$:
\begin{equation}
Z_q [g,\theta ;\mathcal{B}=0]
=\frac{1}{q}\sum_{k=0}^{q-1}\int \Diff \mathcal{B} 
  Z_q[g,\theta ;\mathcal{B}]\rme^{\im k\int \mathcal{B}} 
= \frac{1}{q} \sum_{k=0}^{q-1}
Z_1 \left[ qg, \frac{\theta +2\pi k}{q}   ;\mathcal{B}=0 \right] .
\label{eq:decomposition}
\end{equation}
Therefore, the charge-$q$ Schwinger model can be understood as the disjoint union of the charge-$1$ Schwinger models with different vacuum angles. 
In general, such an operation could have violated the locality property of QFTs, but the charge-$q$ Schwinger model is well-defined as a local QFT~\cite{Pantev:2005rh, Pantev:2005wj, Pantev:2005zs, Hellerman:2006zs, Hellerman:2010fv}. 

This feature of the charge-$q$ Schwinger model is called the decomposition of QFT in Refs.~\cite{Pantev:2005rh, Pantev:2005wj, Pantev:2005zs, Hellerman:2006zs, Hellerman:2010fv} as the Hilbert space on the closed space completely decomposes into $q$ distinct sectors. 
Each sector of the decomposition is called as the universe~\cite{Hellerman:2006zs, Tanizaki:2019rbk, Komargodski:2020mxz, Cherman:2020cvw}. 
The above relation \eqref{eq:decomposition} also tells us that
changing $\theta\rightarrow \theta +2\pi$ gives 
a jump of a universe to a next universe and
each universe has the periodicity $\theta \sim \theta +2\pi q$
while the whole charge-$q$ Schwinger model has 
the $\theta$-periodicity $\theta \sim \theta +2\pi $.
If we put the theory on space with boundary, then
we automatically pick up one specific universe and 
no longer have the sum over the universes.
Although the $1$-form symmetries in higher-dimensions do not have the above decomposition feature, 
we can still generalize the notion of the decomposition to $d$-dimensional QFTs with $(d-1)$-form symmetries~\cite{Tanizaki:2019rbk}.

\subsection{Chiral symmetry and 't Hooft anomaly}

When we set the fermion mass to be zero, $m=0$, the classical theory enjoys the $U(1)$ axial symmetry, 
\begin{equation}
    \overline{\psi}\mapsto \overline{\psi}'=\overline{\psi} \rme^{-\im \alpha \overline{\gamma}_{\mathrm{E}}}, \quad 
    \psi \mapsto \psi'=\rme^{-\im \alpha \overline{\gamma}_{\mathrm{E}}}\psi, 
\end{equation}
where $\overline{\gamma}_{\mathrm{E}}=\im \gamma^{1}_{\mathrm{E}}\gamma^{2}_{\mathrm{E}}$ gives the chirality. 
Quantum mechanically, we have the following change of the path integral measure \cite{Fujikawa:1979ay}
\begin{equation}
    \Diff \overline{\psi}'\Diff \psi'=\Diff \overline{\psi}\Diff \psi \exp\left(\frac{2\im q \alpha}{2\pi}\int \diff A\right), 
\end{equation}
and thus the continuous axial symmetry is explicitly broken.
This is called the Adler-Bell-Jackiw (ABJ) anomaly~\cite{Adler:1969gk, Bell:1969ts}. 
When $q>1$, however, there still exists a nontrivial subgroup, $\mathbb{Z}_{2q}$, of the axial transformation, which gives the genuine symmetry, 
\begin{equation}
     \overline{\psi}\mapsto \overline{\psi} \rme^{-\im \frac{2\pi}{2q} \overline{\gamma}_{\mathrm{E}}}, \quad 
    \psi'=\rme^{-\im \frac{2\pi}{2q} \overline{\gamma}_{\mathrm{E}}}\psi. 
\end{equation}
We note that $\mathbb{Z}_2\subset \mathbb{Z}_{2q}$ gives the fermion parity, which is a part of the $U(1)$ gauge redundancy, and thus the axial symmetry group is given by $(\mathbb{Z}_{q})_\chi$. 

Under the presence of the background gauge field for the $1$-form symmetry, the discrete chiral symmetry is broken:
\begin{equation}
    (\mathbb{Z}_q)_\chi: Z_q[g,\theta ;\mathcal{B}] \mapsto \rme^{\im \int \mathcal{B}} Z_q[g,\theta ;\mathcal{B}] . 
\label{eq:anomaly_Schwinger}
\end{equation}
This is the mixed 't~Hooft anomaly between the $1$-form symmetry and the discrete chiral symmetry~\cite{Anber:2018jdf, Anber:2018xek, Armoni:2018bga, Misumi:2019dwq}. 
The anomaly matching condition claims that there must be $q$ vacua associated with the spontaneous chiral symmetry breaking. 

It is convenient to rephrase this fact from the viewpoint of (extended) operators. 
In $2$ spacetime dimensions, the $0$-form (ordinary) symmetries are generated by the topological $1$-dimensional objects, 
and the $1$-form symmetries are generated by the topological $0$-dimensional objects,
i.e.~local operators with the scaling dimension $=0$. 
In the charge-$q$ massless Schwinger model, there has to be a line operator $U_\chi(L)$ defined on the loop $L$, which generates the $\mathbb{Z}_q$ chiral symmetry, and there is also a local operator $V(x)$, which generates the $1$-form symmetry. 
These operators are topological in the sense that infinitesimal deformations do not affect the correlation functions, which generalizes the Ward-Takahashi identity. 

The fact that $U_\chi(L)$ generates the $\mathbb{Z}_q$ chiral symmetry can be seen from its commutation relation with the chiral condensate operator, 
\begin{\eq}
O_\pm(x)
:=\overline{\psi}(1\pm  \overline{\gamma}_\mathrm{E})\psi(x) .
\end{\eq}
Its real and imaginary parts are called the scalar and pseudo-scalar condensates, respectively:
\begin{equation}
    O_{\pm}(x)=S(x)\pm \im\, P(x). 
\label{eq:def-chiral}
\end{equation}
Let us consider the situation, where $x$ is inside the loop $L$ and we deform $L$ to $L'$ so that $x$ is outside the loop $L'$, and then  
\begin{equation}
    U_\chi(L)O_{\pm}(x)= \rme^{\pm \frac{2\pi \im}{q}}U_\chi(L') O_{\pm}(x). 
    \label{eq:chiral_symmetry}
\end{equation}
This equality is understood as the operator identity, so we assume that we can deform the loop $L$ to $L'$ without crossing any operators except $O_\pm(x)$. 

Similarly, the presence of the $\mathbb{Z}_q^{[1]}$ symmetry implies the presence of the topological point-like operator $V(x)$, which has the nontrivial commutation relation with the Wilson loop $W(C)=\exp(\im \oint_C A)$. 
Let us move the location of $V$ from a point $x$ of the inside of $C$ to another point $x'$ of the outside of $C$, then 
\begin{equation}
    V(x) W(C)=\rme^{\frac{2\pi \im }{q}}V(x') W(C). 
    \label{eq:1form_symmetry}
\end{equation}

So far, everything we mentioned about $U_\chi(L)$ and $V(x)$ is the defining property of symmetry generators, and there is nothing specific to the charge-$q$ Schwinger model. 
The interesting feature of the charge-$q$ Schwinger model is the mixed 't~Hooft anomaly~\eqref{eq:anomaly_Schwinger}. 
In the language of the symmetry generators, the 't~Hooft anomaly implies the nontrivial commutation relation between the symmetry generators,
\begin{equation}
    U_\chi(L) V(x) = \rme^{\frac{2\pi \im}{q}} U_\chi(L') V(x'), 
    \label{eq:anomaly_generator}
\end{equation}
where $x$ is inside of $L$ and $x'$ is the outside of $L'$. 

Let us compare \eqref{eq:anomaly_generator} with \eqref{eq:chiral_symmetry}. 
This suggests that the chiral condensate operator $O_+(x)$ becomes with the $1$-form symmetry generator $V(x)$ by performing the renormalization group (RG) transformation.\footnote{To be precise, we assume the existence of the mass gap for this statement, and we know that the massless Schwinger model is gapped due to the ABJ anomaly. } 
Similarly, the comparison between \eqref{eq:anomaly_generator}  and \eqref{eq:1form_symmetry} shows that $W(C)$ is identified with $U_\chi(C)$ via the RG transformation. 
Therefore, we can confirm the existence of 't~Hooft anomaly by checking the following equality, 
\begin{equation}
    \langle W(C) O_+(x)\rangle = \rme^{\frac{2\pi \im}{q}} \langle W(C) O_+(x')\rangle, 
    \label{eq:relation_Wilson_chiral}
\end{equation}
where $x$ is inside of $C$, $x'$ is outside of $C$, and $x, x'$ are sufficiently far away from the loop $C$. 
For this condition to be satisfied, we must make the loop $C$ sufficiently larger than the size of mass gap.  

\subsection{Analytical results}
So far, we have discussed the kinematical aspects of the charge-$q$ Schwinger model.
As the massless Schwinger model is exactly solvable~\cite{Schwinger:1962tp, Lowenstein:1971fc, Coleman:1975pw, Coleman:1976uz, Manton:1985jm, Faddeev:1986pc, Jayewardena:1988td, Hetrick:1988yg, Iso:1988zi, Smilga:1992hx, Adam:1993fc, Adam:1997wt}, we can confirm these features by an explicit calculation. 

It is known that
there is a correspondence between Dirac fermion and compact boson in two dimensions.
This is called Abelian bosonization.
Specifically, 
the kinetic term $\overline{\psi}\slashed{\partial}\psi$ for the Dirac fermion 
is mapped to $\frac{1}{8\pi}|\diff \phi|^2$ 
with a $2\pi$-periodic scalar field $\phi$ . 
The vector and axial currents correspond to 
\begin{equation}
    \overline{\psi} \gamma_{\mathrm{E}}^\mu \psi \leftrightarrow \frac{1}{2\pi}\varepsilon^{\mu\nu}\partial_\nu \phi, 
    \quad \overline{\psi}\, \overline{\gamma}_\mathrm{E} \gamma_{\mathrm{E}}^\mu \psi \leftrightarrow \frac{1}{2\pi}\partial_\mu \phi.
\end{equation}
The chiral condensate operators are related as 
\begin{equation}
    O_{\pm}(x) \leftrightarrow C(M)\rme^{\pm \im \phi}, 
\end{equation}
where $C(M)$ is a multiplicative renormalization constant proportional to the renormalization scale $M$. 

The bosonized action of the charge-$q$ Schwinger model is given by 
\begin{equation}
    S
    =\int_{M_2}\left(\frac{1}{2g^2}\diff A \wedge \star\diff A
    +\frac{\im}{2\pi}(q\:\! \phi-\theta) \diff A
    + \frac{1}{8\pi} d\phi \wedge \star \diff \phi \right). 
\end{equation}
In the bosonized description, the discrete chiral symmetry is realized as the shift symmetry, $\phi\mapsto \phi+\frac{2\pi}{q}$. 
By completing the square in terms of $\diff A$, we find that $\phi$ becomes the massive boson, 
\begin{equation}
    S
=\frac{1}{8\pi} \int_{M_2}\left((\partial_\mu\phi)^2+\frac{ g^2}{\pi}(q\:\!\phi-\theta)^2\right)\diff^2 x
+\frac{1}{2g^2} \int_{M_2} \left(F_{01}+\frac{\im g^2}{2\pi}(q\:\!\phi-\theta)\right)^2\diff^2 x. 
\end{equation}
Although this expression does not respect the $2\pi$ periodicity of the compact boson $\phi$, it is useful to identify the mass gap $\mu^2$,
\begin{equation}
    \mu^2=\frac{q^2 g^2}{\pi}. \label{eq:mass-gap}
\end{equation}
Since $\phi$ and $A$ are dual to each other via the topological coupling, this is often called as the photon mass of the massless Schwinger model. 

In order to understand the consequence of 't~Hooft anomaly, we are interested in the phase of the chiral condensate operator. 
For this purpose, we can take the massive limit, $g\to \infty$, where the mass gap becomes infinite. 
In this case, the path integral for $A$ becomes 
\begin{equation}
    \int \Diff A \exp\left(\frac{\im }{2\pi}
    \int_{M_2} (\theta -q\:\! \phi)
    \diff A\right)\propto \delta(\diff \phi)\sum_{n\in \mathbb{Z}}\rme^{\im n (\theta-q \phi)}. 
\end{equation}
The first factor on the right-hand-side is the consequence of the equation of motion, $\diff \phi=0$, and thus $\phi$ should be constant to obtain the nontrivial values of this path integral. 
The second factor comes from the summation over the instanton sectors, and this becomes nonzero only when $\theta-q\:\!\phi \in 2\pi \mathbb{Z}$. 
Therefore, because of the $2\pi$ periodicity of $\phi$, 
possible values of $\phi$ are $\phi =(\theta +2\pi k) /q$ with $k=0,1,\ldots, q-1$. 
This implies that
the theory has $q$ distinct vacua characterized by
\begin{equation}
    \langle\rme^{\im \phi}\rangle_k 
\sim \rme^{\im \frac{\theta+2\pi k}{q}} .
\end{equation}
This is exactly what we expect 
for the spontaneous breaking of the $\mathbb{Z}_q$ chiral symmetry 
as a consequence of the 't~Hooft anomaly matching. 

In order to confirm \eqref{eq:relation_Wilson_chiral}, we should perform the path integral under the presence of the Wilson loop $W(C)$. 
Since we can write $W(C)^p=\rme^{\im p \int_D \diff A}$ with $C=\partial D$, the Wilson loop can be represented as the spacetime-dependent $\theta$ angle, where 
\begin{equation}
    \theta(x)
    =\begin{cases}
    \theta & {\rm for}\ x\not \in D \\
    \theta+2\pi p & {\rm for}\ x\in D
    \end{cases}  .
\end{equation}
We can perform the path integral of $A$ exactly in the same way, and we find 
\begin{equation}
    \langle \rme^{\im \phi(x)}\rangle_k
\sim \rme^{\im \frac{\theta(x)+2\pi k}{q}}. 
\end{equation}
Since the value of the $\theta$ angle jumps by $2\pi$ across the Wilson loop, we obtain the relation~\eqref{eq:relation_Wilson_chiral} as required by the 't~Hooft anomaly combined with the RG argument. 

While the above discussion gives an immediate confirmation of the relation~\eqref{eq:relation_Wilson_chiral},
we have taken the limit $g\rightarrow \infty$ in which the mass gap is infinite.
Let us also confirm the relation~\eqref{eq:relation_Wilson_chiral} in a more elementary fashion keeping $g$ finite. 
In this case, we set $M_2=\mathbb{R}^2$ and take the rectangular Wilson loop $W(C_{T\times L})^p$, where $C_{T\times L}=\partial([-T/2,T/2]\times [-L/2,L/2])$. 
By taking the limit $T\to \infty$, the classical configuration becomes constant along the imaginary-time direction, and thus the classical equation of motion becomes 
\begin{equation}
    -\partial_x^2 \phi(x) +\mu^2 \left( \phi(x) -\frac{\theta(x)}{q} \right) =0,
\end{equation}
where 
\begin{equation}
    \theta(x)
    =\begin{cases}
    \theta&  {\rm for}\ |x| > \frac{L}{2}  \\ 
    \theta+2\pi p &  {\rm for}\ |x|< \frac{L}{2}  
    \end{cases} .
\end{equation}
We solve this equation with the boundary condition $\phi(x)\to \frac{\theta}{q}$ for $x\to \pm \infty$. We then obtain 
\begin{equation}
    \phi(x)
    =\frac{\theta}{q}+
    \begin{cases}
    \frac{2\pi p}{q} \sinh\left(\frac{\mu L}{2}\right)\rme^{\mu x} & {\rm for}\ x<-\frac{L}{2} \\
    \frac{2\pi p}{q}\left(1-\rme^{-\frac{\mu L}{2}} \cosh \left(\mu x\right)\right) & {\rm for}\ -\frac{L}{2}<x<\frac{L}{2}\\
    \frac{2\pi p}{q} \sinh\left(\frac{\mu L}{2}\right)\rme^{-\mu x} & {\rm for}\ x>\frac{L}{2}
    \end{cases} .
    \label{eq:phase_rotation}
\end{equation}
This tells us that 
$\phi (x)$ is almost constant away from the Wilson loop at $x=\pm L/2$:
\begin{equation}
\phi(x)
\simeq  \begin{cases}
    \frac{\theta}{q} & {\rm for}\ |x| \gg \frac{L}{2} \\
    \frac{\theta +2\pi p}{q} & {\rm for}\  |x| \ll \frac{L}{2}
    \end{cases} .
\end{equation}
In particular, when $\mu L$ is large,
$\phi (x)$ around the Wilson loop quickly changes from $\theta /q$ to $(\theta +2\pi p)/q$ 
if we go from outside to inside.
In terms of $\phi (x)$, the chiral condensate operator under the presence of the Wilson loop is then given by 
\begin{equation}
O_+ (x)
=S(x)+\im\:\! P(x)
=\frac{\rme^\gamma q g}{2\pi^{3/2}}\rme^{\im \phi(x)}, 
    \label{eq:condensate_with_wilsonloop}
\end{equation}
where the overall coefficient is determined from the knowledge on the charge-$1$ Schwinger model~\cite{Jayewardena:1988td, Hetrick:1988yg, Iso:1988zi, Smilga:1992hx, Adam:1993fc, Adam:1997wt}. 
Therefore, this is also almost constant away from the Wilson loop:
\begin{equation}
S(x)+\im\:\! P(x)
\simeq  \begin{cases}
    \frac{\rme^\gamma q g}{2\pi^{3/2}}\ \rme^{\im \frac{\theta}{q}} & {\rm for}\ |x| \gg \frac{L}{2} \\
    \frac{\rme^\gamma q g}{2\pi^{3/2}}\ \rme^{\im \frac{\theta +2\pi p}{q}} & {\rm for}\  |x| \ll \frac{L}{2}
    \end{cases} ,
\label{eq:condensates_anomaly}
\end{equation}
where the ${\rm mod}(q)$ structure of the $p$-dependence reflects the $\mathbb{Z}_q$ 1-form symmetry.
When we cross the Wilson loop from outside to inside, 
it quickly rotates along the circular arc from the angle $\theta /q$ to $(\theta +2\pi p)/q$,
where rotating direction is counterclockwise for $p>0$ and clockwise for $p<0$.

\begin{figure}[tbp]
    \centering
    \includegraphics[scale=0.6]{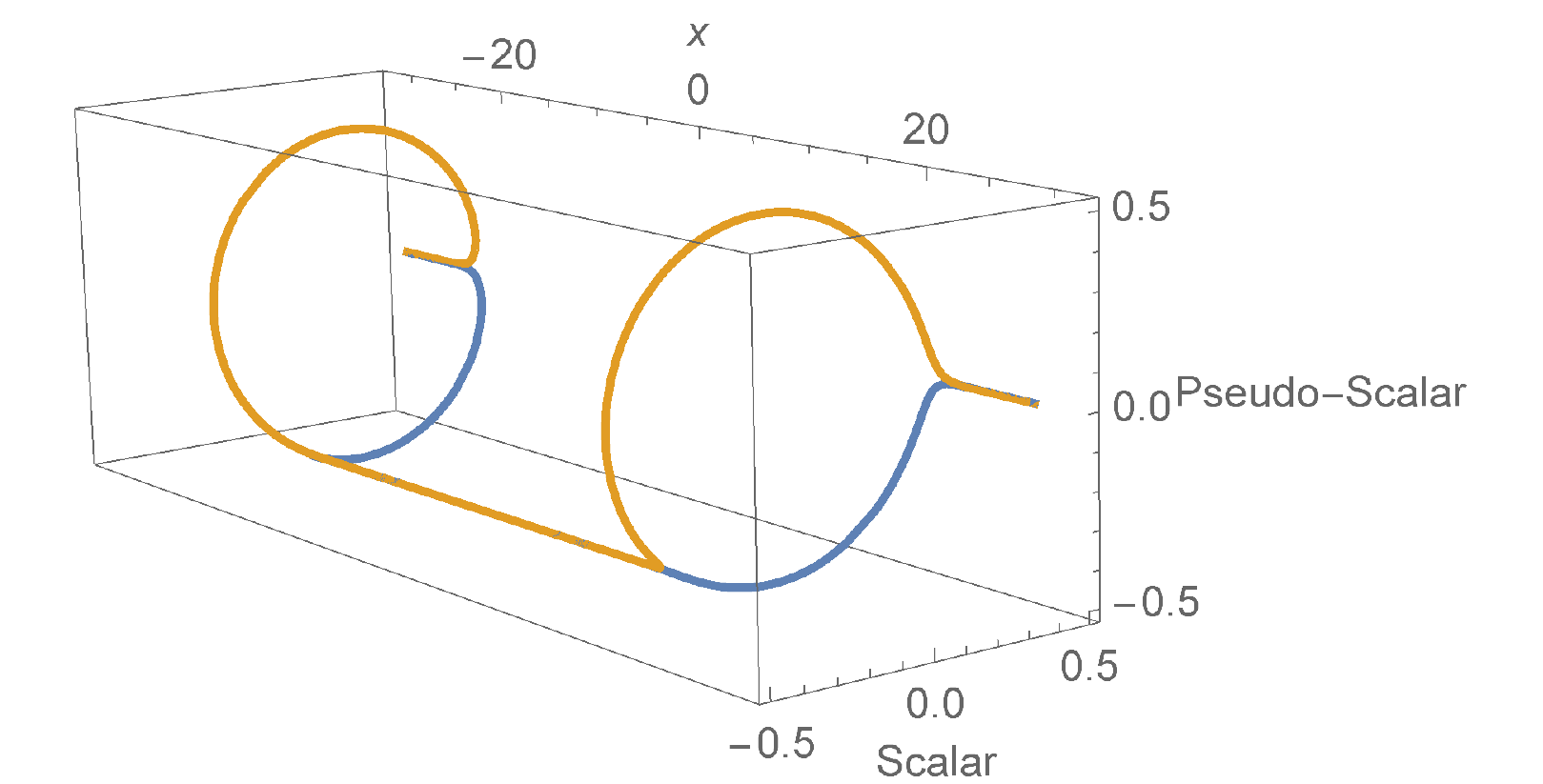}
    \caption{
    Position dependence of the scalar condensate $S(x)$ and pseudo-scalar condensate $P(x)$ in the massless charge-$3$ Schwinger model with the charge-$q_p$ Wilson loop.
    We set $g=1$, $\theta =0$ and put the Wilson loop at $x=\pm 20$ (i.e.~$L=40$). 
    The yellow and blue curves denote the cases for the probe charges $q_p=2$ and $q_p=-1$, respectively.
    }
    \label{fig:condensates_3d}
\end{figure}

Figure~\ref{fig:condensates_3d} illustrates the behaviors of $S(x)$ and $P(x)$, given by \eqref{eq:condensate_with_wilsonloop} with the $x$-dependent phase factor \eqref{eq:phase_rotation}.
The probe charge is put at $x=\pm 20$, i.e. $L=40$ and it is sufficiently large compared with the mass gap $\mu=\frac{q g}{\sqrt{\pi}}\simeq 1.7$. 
We can see that the chiral condensate rotates its phase by $120$ degrees ($2\pi/3$ in radian)
when crossing the Wilson loop $W^{q_p}$ with $q_p=\pm 1 \bmod 3$.
This is nothing but the anomaly relation~\eqref{eq:relation_Wilson_chiral}.

\section{Lattice Hamiltonian formulation and numerical setup}
\label{sec:lattice}

In this section, we describe the lattice Hamiltonian formulation of the charge-$q$ Schwinger model and explain our setup for the numerical computations. 
In Sec.~\ref{sec:Hamiltonian}, we introduce the lattice Hamiltonian $U(1)$ gauge theory with the staggered fermion~\cite{Banks:1975gq, Carroll:1975gb} with a special emphasis on the charge-$q$ model~\cite{Honda:2021ovk}. 
We here adopt the recent proposal~\cite{Dempsey:2022nys} for the correspondence between the lattice and continuum parameters. 
In Sec.~\ref{sec:observables}, 
we define local operators such as energy density and chiral condensate operators, 
and their UV renormalization shall be discussed in detail. 
In Sec.~\ref{sec:simulation_method}, we explain the simulation method. 

\subsection{Lattice Hamiltonian for  charge-\texorpdfstring{$q$}{q} Schwinger model with \texorpdfstring{$\theta$}{theta} term}
\label{sec:Hamiltonian}

The continuum Lagrangian in the Minkowski metric $\eta_{\mu\nu}=\mathrm{diag}(1,-1)$ is given by
\be
\mathcal{L}
=\frac{1}{2g^2}F_{01}^2 +\frac{\theta}{2\pi}F_{01}
+\im\, \overline{\psi} \gamma^\mu (\p_\mu + \im\:\! q\:\! A_{\mu})\psi-m\,\overline{\psi}\psi,
\ee
where we take the gamma matrices as $\gamma^0 =\sigma^3$, $\gamma^1 =\im \sigma^2$ and $\overline{\gamma} =\gamma^0\gamma^1=\sigma^1$.
We take the temporal gauge $A_0=0$ for the canonical quantization. 
The Gauss law is obtained by the equation of motion for $A_0$,
\be
\p_1 \left(\frac{1}{g^2} F_{01}+\frac{\theta}{2\pi}\right)= q \, \psi^\dagger \psi(x), 
\label{eq:Gauss_law_classical}
\ee
where the dynamical charge $q$ appears explicitly on the right-hand-side. 
Introducing the canonical momentum $\Pi(x)=\frac{\delta L}{ \delta \dot{A_1}(x)}$, 
the Hamiltonian is given by
\begin{\eq}
H(x)
=\frac{g^2}{2} \left( \Pi -\frac{\theta}{2\pi} \right)^2
 -\im\, \bar{\psi} \gamma^1 (\del_1 +\im\:\! q\:\! A_1) \psi +m\bar{\psi} \psi .
\label{eq:hamiltonian-cont}
\end{\eq}
The Gauss law constraint (\ref{eq:Gauss_law_classical}) is rewritten as 
\be 
\p_1 \Pi(x)=q \, \psi^\dagger \psi(x).
\label{eq:Gauss_continuum}
\ee

The lattice regularization of the model can be found as follows.
Introducing a lattice with $N$ sites and lattice spacing $a$, we define the staggered fermion $\chi_n$ for the two-component Dirac fermion $\psi (x)$, where $n$ labels the lattice site ($x=na$). 
Here, $\chi_n$ is a single component complex fermionic operator, and the Dirac fermion at $x$ extends over two sites on the lattice:
\begin{equation}
\psi(x) \leftrightarrow
\frac{1}{\sqrt{a}} 
\begin{pmatrix}    
\chi_{2\lfloor n/2\rfloor} \cr
\chi_{2\lfloor n/2\rfloor+1}
\end{pmatrix}. 
\label{eq:fermion_lattice}
\end{equation}
The gauge field and its canonical momentum are represented by the link variables,
\begin{\eq}
U_n \leftrightarrow \rme^{\im a  A_1 (x)} ,\quad L_n \leftrightarrow -\Pi(x),		
\end{\eq}
which are defined on the link between the sites $n$ and $n+1$.
The canonical commutation relations for the lattice fields are given
\begin{\eq}
[L_n, U_m] = U_m\delta_{nm} ,\quad \{ \chi_n ,\chi_m^\dag \} = \delta_{nm}.
\label{eq:commutation_relation}
\end{\eq}
Then the lattice Hamiltonian is 
\begin{\eqa}
H 
= J \sum_{n=0}^{N-2} \left( L_n +\frac{\theta}{2\pi}  \right)^2 
 -\im w \sum_{n=0}^{N-2} \Bigl[ \chi_n^\dag (U_n)^q \chi_{n+1} -{\rm h.c.} \Bigr]  
 -m_{\mathrm{lat}}\sum_{n=0}^{N-1} (-1)^n \chi_n^\dag \chi_n ,
\label{eq:lattice_Hamiltonian}
\end{\eqa}
where 
\begin{\eq}
J=\frac{g^2 a}{2} ,\quad  w=\frac{1}{2a} .
\end{\eq}
We relate the lattice fermion mass $\mlat$ with the continuum fermion mass $m$ as 
\begin{equation}
    m_{\mathrm{lat}}=m-\frac{q^2 g^2 a}{8}, 
    \label{eq:lattice_cont_corr}
\end{equation}
following the recent proposal~\cite{Dempsey:2022nys}, which is reviewed in Appendix~\ref{app:discreteanomaly_lattice}.

As in the continuum theory,
physical states are constrained by the lattice version of the Gauss law
\begin{\eq}
L_n -L_{n-1} 
= q \Biggl[ \chi_n^\dag \chi_n -\frac{1-(-1)^n}{2}  \Biggr].
\label{eq:Gauss_law}
\end{\eq}
We remove all degree of freedom for the variables $(U_n ,L_n )$ by taking open boundary condition $L_0 =0$, solving the Gauss law and fixing the gauge so that $U_n=1$ for all $n$.
Then the Hamiltonian is written pure in terms of the fermion operators as
\begin{align}
H 
=& -\im w \sum_{n=0}^{N-2} \Bigl[ \chi_n^\dag  \chi_{n+1} -{\rm h.c.} \Bigr] 
 -\mlat \sum_{n=0}^{N-1} (-1)^n \chi_n^\dag \chi_n  \notag\\
 & +J \sum_{n=0}^{N-2} \Biggl[\frac{\theta}{2\pi} +q \sum_{j =0}^n \left( \chi_j^\dag \chi_j -\frac{1-(-1)^j}{2} \right)  \Biggr]^2 ,
\label{eq:lattice_hamiltonian_0}
\end{align}
which acts on a finite dimensional Hilbert space.
We note that the periodicity of $\theta$ is lost by taking the open boundary condition in this formulation.

To see the 't Hooft anomaly, we introduce the Wilson loop, which size is $\hat{\ell}$, on the lattice.
It corresponds to putting the two probe charges $+q_p$ and  $-q_p$ with a distance $\hat{\ell}$.
It is realized by 
introducing the position-dependent $\theta$-angle,
\begin{align}
\label{eq:def-vartheta}
\vartheta_n 
= \left\{
 \begin{array}{ccc}
  \theta +2\pi q_p  & & \ {\rm for}\ \hat{\ell}_0 \le n < \hat{\ell}_0 +\hat{\ell}, \\
  \theta & & \text{otherwise}. 
 \end{array} 
 \right.
\end{align}
Under the open boundary condition, it would be appropriate to take
\begin{\eq}
\hat{\ell}_0 = \frac{N-\hat{\ell} -1}{2}  ,
\end{\eq}
with odd $\hat{\ell}$ for even $N$ and even $\hat{\ell}$ for odd $N$. 
The lattice Hamiltonian in the presence of the probes is
\begin{align}
H 
= & -\im w \sum_{n=0}^{N-2} \Bigl[ \chi_n^\dag  \chi_{n+1} -{\rm h.c.} \Bigr] 
 -\mlat \sum_{n=0}^{N-1} (-1)^n \chi_n^\dag \chi_n  \notag\\
 & +J \sum_{n=0}^{N-2} \Biggl[\frac{\vartheta_n }{2\pi} +q \sum_{j =0}^n \left( \chi_j^\dag \chi_j -\frac{1-(-1)^j}{2} \right)  \Biggr]^2 ,\label{eq:Hamiltonian-final}
\end{align}
where $\theta$ in \eqref{eq:lattice_hamiltonian_0} is replaced by the position dependent one, $\vartheta_n$ in \eqref{eq:def-vartheta}.

At the end of this subsection, let us give a relation between $q=1$ and general charge-$q$ Schwinger model:
\begin{\eq}
H \left( q , J ,  \vartheta_n \right)
= H\left( 1 , q^2 J ,  \frac{\vartheta_n}{q} \right), 
\label{eq:relation-q-Schwinger}
\end{\eq}
and this realizes the decomposition~\eqref{eq:decomposition} in the case of the open boundary condition. 
This translation to $q=1$ is possible because we take the open boundary condition. 
If we took the periodic boundary condition, then we could not eliminate the spatial link variables completely by gauge fixing. 
The spatial hopping term, $\chi^\dagger_n(U_n)^q\chi_{n+1}$, genuinely depends on the choice of $q\ge 1$, and we cannot relate them by simple replacements of coupling constants.

\subsection{Local observables and UV renormalization}
\label{sec:observables}

In this subsection, we define energy density and chiral condensates as local observables in the lattice Hamiltonian formalism and discuss the treatment of their UV divergence. 
Local operators play the central role in quantum field theories. 
Moreover, when we use the open boundary condition, use of local operators has a huge advantage to reduce the boundary effects~\cite{Honda:2021ovk}.

Let us first define the local behavior of the energy $E(n)$ at each site $n$. 
As it is defined by the Hamiltonian density, we can simply obtain it by extracting the summand of \eqref{eq:Hamiltonian-final} with one caution. 
Since the staggered fermion is used, a pair of even and odd sites forms the actual spatial point and thus we have to take an average between the neighboring sites to define the physical local quantities. 
As a result, we define the site-dependent energy density by\footnote{In this paper, we take a specific averaging around the site $n$. We may also consider more smooth one, such as $\sum_{n'}\mathcal{N}\exp\left(-\frac{(n-n')^2}{2n_0^2}\right)h^{w,M,J}_{n'}$, and then the energy density is averaged over the region $\Delta x=a n_0$. We can regard it as a local operator as long as it satisfies $a\ll an_0\ll \xi=\mu^{-1}$.} 
\begin{equation}
E_{\mathrm{bare}}(n) = \underbrace{\left( \frac{h_{n-1}^{w}}{4} + \frac{h_n^{w}}{2} + \frac{h_{n+1}^{w}}{4}  \right)}_{\mbox{fermion kinetic term}} + \underbrace{\left( \frac{h_{n-1}^M}{4} + \frac{h_n^M}{2} + \frac{h^M_{n+1}}{4} \right)}_{\mbox{fermion mass term}} +\underbrace{\left( \frac{h_{n-1}^{J}}{4} + \frac{h_n^{J}}{2} + \frac{h_{n+1}^{J}}{4} \right)}_{\mbox{gauge kinetic term}},
\label{eq:def-local-energy}
\end{equation}
where 
\beq
&& h_n^{w} = -\im w (\chi_n^\dag  \chi_{n+1} - \chi_{n+1}^\dag \chi_n ), \quad
h_n^M = -(-1)^n m_{\rm lat} \:\! \chi_n^\dag \chi_n ,   \nonumber\\
&& h_n^J = J  \Biggl[\frac{\vartheta_n }{2\pi} +q \sum_{j =0}^n \left( \chi_j^\dag \chi_j -\frac{1-(-1)^j}{2} \right)  \Biggr]^2 .
\eeq
This is a bare quantity, and its expectation value $\langle E_{\mathrm{bare}}(n)\rangle$ is UV divergent when we take $a\to 0$. 
We note that the Schwinger model is super-renormalizable as it only has the dimensionful couplings, so its UV divergence can be subtracted by the normal-ordering procedure. 
Furthermore, since the UV divergence does not depend on the $\theta$ parameter in our formulation, it can be simply achieved by subtracting the ground-state expectation value at $\theta=0$:
\begin{equation}
    E(n)=E_{\mathrm{bare}}(n)
    -\left. \left\langle E_{\mathrm{bare}}\left( [ N/2 ] \right) \right\rangle \right|_{\theta=0}. 
\end{equation}

Next, we define the local scalar condensate, $\overline{\psi}\psi \rightarrow S(n)$, and local pseudo-scalar condensate, $ -\im \overline{\psi} \overline{\gamma} \psi \rightarrow P(n)$, at site $n$ by using the correspondence \eqref{eq:fermion_lattice}:
\beq
S_{\mathrm{bare}}(n) &=& \frac{s_{n-1}}{4}+ \frac{s_n}{2}+ \frac{s_{n+1}}{4}, \quad P_{\mathrm{bare}}(n) = \frac{p_{n-1}}{2}+\frac{p_{n}}{2}, 
\eeq
where
\beq
s_n = (-1)^n w \,\chi_n^\dag \chi_n , \quad 
p_{n}= (-1)^{n+1}  w \:\! (\chi_n^\dag \chi_{n+1} -\chi_{n+1}^\dag \chi_n ).
\eeq
Here, we take suitable average over neighboring sites to take into account the even-odd inequality of staggered fermion.
As we have done for the energy density, one can eliminate UV divergences of these quantities by subtracting their expectation values at $\theta=0$. 
However, as we are going to confirm the anomaly relation~\eqref{eq:condensates_anomaly}, we would like to define the origin of the chiral condensate in the limit $m\to 0$. 

As the UV divergence of chiral condensates comes only from the fermion one-loop diagram due to the super-renormalizability, we can evaluate the UV-divergent piece using free fermion:
\begin{align}
S_{\mathrm{div}}
&=-\frac{m}{\pi \sqrt{1+(\mu a)^2}}\, K\left(\frac{1}{\sqrt{1+(\mu a)^2}}\right)\nonumber\\
&=-\frac{m_{\mathrm{lat}}+\frac{q^2 g^2 a}{8}}{\pi \sqrt{1+(\mu a)^2}}\, K\left(\frac{1}{\sqrt{1+(\mu a)^2}}\right),
\label{eq:S-tree}
\end{align}
where $K(k)$ denotes the complete elliptic integral of the first kind,
\beq
K(k)=\int_0^{\pi/2} \frac{1}{\sqrt{1-k^2 \sin^2 t}} \diff t. 
\eeq
We have replaced the fermion mass parameter in the loop integral by the mass gap $\mu$ as it does not change the UV structure and it circumvents the IR singularity in the chiral limit. 
The renormalized condensates are given by 
\begin{equation}
    S(n)=S_{\mathrm{bare}}(n)-S_{\mathrm{div}},\quad
    P(n)=P_{\mathrm{bare}}(n). 
\end{equation}
When we subtract the UV divergence $S_{\mathrm{div}}$, the $\mathcal{O}(a)$ correction~\eqref{eq:lattice_cont_corr} for one-dimensional staggered fermion proposed in Ref.~\cite{Dempsey:2022nys} is taken into account. 
As continuum fermion mass and lattice fermion mass are different, we have a nonzero subtraction even for $m_{\mathrm{lat}}=0$. 
As $a\to 0$, $S_{\mathrm{div}}$ behaves as 
\begin{equation}
    S_{\mathrm{div}}=\left(\frac{m_{\mathrm{lat}}}{\pi}+\frac{q^2 g^2 a}{8\pi}\right)\left(\ln \frac{\mu a}{4}+\mathcal{O}((\mu a)^2 \ln (\mu a))\right). 
\end{equation}
We note that there is no UV divergence if we set $m_{\mathrm{lat}}=0$ and this is consistent with the super-renormalizability. However, the convergence in the limit $a\to 0$ behaves as $\mathcal{O}(a\ln a)$ and this is slower than polynomials.\footnote{The presence of $\mathcal{O}(a\ln a)$ in the continuum limit itself has been recognized, for example, in Ref.~\cite{Banuls:2016lkq}. 
However, to our best knowledge, its physical origin has not been known so its coefficient is treated as one of fitting parameters. We here show that it comes from the $\mathcal{O}(a)$ shift for the lattice mass parameter, and its coefficient is fixed by the requirement of the discrete axial anomaly relation. } 
Subtraction of $S_{\mathrm{div}}$ is practically useful to achieve the continuum limit of the scalar condensate\footnote{
Another option is to set $m_{\mathrm{lat}}=-\frac{q^2 g^2 a}{8}$ to achieve the chiral limit, $m=0$, and then we need no subtractions as $S_{\mathrm{div}}=0$. This is the procedure suggested by Ref.~\cite{Dempsey:2022nys} and it is indeed more convenient than setting $m_{\mathrm{lat}}=0$. 
In this paper, however, we do not take this option and we simply set $m_{\mathrm{lat}}=0$ to study the chiral limit, since this is the numerical setup in almost all previous literature and it would be easier for readers to compare our results with them. } 
because we can take the polynomial ansatz for the continuum extrapolation after the subtraction of $S_{\mathrm{div}}$.

\subsection{Simulation method}
\label{sec:simulation_method}

In our previous papers~\cite{Honda:2021aum, Honda:2021ovk}, we employed the adiabatic state preparation formulated for digital quantum simulation to study the ground state of the same model. 
We carried out the numerical simulations up to the lattice size $N=25$ with the IBM qiskit simulator.
In the present paper, to investigate the model more quantitatively,
we scale up the simulation up to the lattice size $N=801$ by applying the density-matrix renormalization group (DMRG) with the ITensor Library~\cite{itensor}.

Here, we first convert the fermionic degrees of freedom into the spin degrees of freedom via the Jordan-Wigner transformation \cite{Jordan1928},
\begin{align}
\chi_n =  \frac{X_n -\im\:\! Y_n}{2} \left( \prod_{i=0}^{n-1} -\im Z_i \right),~~
\chi_n^\dag =  \frac{X_n +\im\:\! Y_n}{2} \left( \prod_{i=0}^{n-1} \im Z_i \right) ,
\end{align}
where ($X_n,Y_n,Z_n$) stands for the Pauli matrices $(\sigma_1 ,\sigma_2 ,\sigma_3 )$ located at site $n$.
Accordingly, the Hamiltonian takes the form, 
\begin{align}
H =&\,  J\sum_{n=0}^{N-2} \left[ q \sum_{i=0}^{n}\frac{Z_i + (-1)^i}{2} +\frac{\vartheta_n}{2\pi}\right]^2 
\nonumber\\ 
&+ \frac{w}{2}\sum_{n=0}^{N-2}\big[X_n X_{n+1}+Y_{n}Y_{n+1}\big]
- \frac{m_{\rm lat}}{2}\sum_{n=0}^{N-1}(-1)^n Z_n , 
\end{align}
and it acts on the Hilbert space $\mathcal{H}\simeq \otimes_{i=1}^{N}\mathbb{C}^2$. 
We represent the wave function $|\Psi\rangle\in \mathcal{H}$ in the form of the matrix-product state (MPS):
\begin{align}
    \ket{\Psi}
    =\sum_{i_1,\dots i_N=1}^{d}\Tr\left(A_1^{(i_1)}A_2^{(i_2)}\cdots A_N^{(i_N)}\right)\ket{i_1 i_2\dots i_{N-1}i_N}.\label{eq:MPS-ansatz}
\end{align}
where $d$ denotes the dimension of local Hilbert space corresponding to a local spin degrees of freedom i.e.~$d=2$ here, and $A_n^{(i_n)}$ is a $D\times D$ complex matrix. 
$D$ is called the bond dimension, and MPS can represent any elements of the Hilbert space when $D\ge d^{N/2}$. 
When applying the DMRG, we assume that the ground-state wave function can be approximated by the MPS with a fixed bond dimension, and the ground state is searched using variational methods within the ansatz. 
The number of variational parameters is roughly given by $N d D^2$, and thus the numerical cost only grows linearly in terms of the system size $N$ when $D$ is treated as a constant. 
Once the ground state is obtained in the form of MPS, the expectation values of local observables can be also efficiently calculated.

This prescription is thought to give a good approximation for searching the ground state of $(1+1)$d quantum many-body systems with local and gapped Hamiltonian. 
The bond dimension gives the upper bound for the entanglement entropy as $S_{\mathrm{EE}}\le \ln D$, and thus DMRG is useful if the entanglement entropy stays constant in the infinite-volume limit. 
As $S_{\mathrm{EE}}$ generically obeys the area law for gapped systems, the above criterion is satisfied for such systems and the DMRG becomes useful for numerical computations.
The Schwinger model in the chiral limit $m\to 0$ is a gapped system, and thus we can expect DMRG is applicable to study the 't~Hooft anomaly of the charge-$q$ Schwinger model.\footnote{
Strictly speaking, the area law of the entanglement entropy for the $(1+1)$d quantum many-body systems is shown only for systems with finite-dimensional local Hilbert space for the gapped Hamiltonian with finite-range interactions~\cite{Eisert:2008ur}. 
We should note that the Schwinger model does not belong to this class: Before solving the Gauss law, the local Hilbert space for the gauge field is infinite-dimensional, and after solving the Gauss law, the range of interaction becomes infinite. 
Still, electric fields of the ground state for the Schwinger model cannot be too large practically; thus, it is reasonable to believe in the validity of the area law. 

Let us give another remark about the bond dimension. The correlation length in the lattice unit is given by $\xi=\frac{1}{\mu a}$, and we can expect that the entanglement entropy behaves as $S_{\mathrm{EE}}=\mathcal{O}(\ln \xi)$. As we approach the continuum limit, the bond dimension should be also increased.  }

\section{Numerical results with density-matrix renormalization group}
\label{sec:results}

In this section, we show the results of DMRG about the charge-$3$ Schwinger model (i.e.~$q=3$)
while we expect similar results for other values of $q >1$. 
The lattice parameters are set to 
$a\in [0.05,\ 0.20]$, $m_{\mathrm{lat}}\in [0,\ 0.60]$ and $\theta\in [0,\ 2\pi q ]$ in the $g=1$ unit. 
The bond dimension $D$ is taken in the range $[200, 300]$ in our simulation. 
We carefully study the $D$-dependence to check the validity of DMRG, and 
we confirm that the observables near the chiral limit ($m_{\mathrm{lat}}\simeq 0$) saturate within 10 digits of precision by taking $D\ge 50$ (see Appendix~\ref{sec:app-D-deps}).

\subsection{Boundary effects and periodicity of theta angle}
Since we are working in the finite size lattice with the open boundary condition,
observables generically receive finite volume corrections including effects of the boundaries.
In particular, not only values of observables deviate from that of the infinite volume limit but also periodicity of theta is broken due to the presence of the boundaries.
Here we study the boundary effects~\cite{Honda:2021ovk}.\footnote{
See also a very recent study \cite{Okuda:2022hsq} of the Schwinger model that relates the boundary conditions between the lattice and continuum theories.
}
We will see that it strongly depends on 
whether or not observables involve operators located close to the boundaries.

Specifically we study the boundary effects 
by comparing behaviors of the local energy $E(n)$ of the ground state
and its average $\varepsilon(\theta)$ over the space:
\begin{\eq}
\varepsilon(\theta) := \frac{1}{L} \sum_{n=0}^{N-1} E(n) ,
\end{\eq}
which is the same as the total ground state energy divided by the physical volume $L$.
In the infinite volume limit,
the translational symmetry is restored and therefore
the both are expected to approach the same value 
unless the site $n$ in the local energy $E(n)$ is close to the boundaries.
Let us denote the above two quantities in the infinite volume and continuum limits for the $k$-th universe by $E_k (\theta )$.
It has been calculated for the $q=1$ case
by the mass perturbation theory\footnote{
The result is given in Eq.~(68) of Ref.~\cite{Adam:1997wt}.
} up to $\mathcal{O}(m^3 )$ \cite{Adam:1997wt}.  
Using the relation \eqref{eq:decomposition}
between the charge-$q$ Schwinger model and $q=1$ case,
one can find the result for generic $q$
by making the replacement $g \rightarrow qg$, $\theta \rightarrow (\theta - 2\pi k )/q$:
\begin{\eq}
E_k  (\theta ) 
=-m\frac{\rme^\gamma qg}{2\pi^{3/2}} \cos{\frac{\theta -2\pi k}{q}}
-m^2 \frac{ \rme^{2\gamma}}{16\pi^2} 
\left( C_+ \cos{\frac{2(\theta -2\pi k )}{q}} +C_- \right)
+\mathcal{O}(m^3 ) ,
\label{eq:Ek}
\end{\eq}
where\footnote{
The precise definitions of $C_+$ and $C_-$ 
(denoted as $\mu^2 E_+ $ and $\mu^2 E_- $ in \cite{Adam:1997wt} respectively) 
are
$C_+ $ $=$ $2\pi \int_0^\infty \diff r \Bigl[ r \left( \rme^{-2K_0 (r) } -1\right) \Bigr] $
and $C_- $ $=$ $4\pi \int_0^\infty \diff r \Bigl[ r \log{r} \left( 
( r K_1 (r) -1 ) \rme^{2K_0 (r) }  +1 \right) \Bigr] $.
} 
$C_+ \simeq -8.9139$ and $C_- \simeq 9.7384$. 
Here we choose the $k=0$ sector by setting $L_0=0$ as the open boundary condition. 
Strictly speaking, $E_k (\theta )$ has a UV divergence dependent on $m$ and
\eqref{eq:Ek} is the expression after a regularization.
Therefore, instead of $E_0 (\theta )$ itself,
we will compare our simulation results with
\begin{align}
    E_0(\theta)-E_0(0)
    = m\frac{\rme^\gamma qg}{2\pi^{3/2}} \left(1-\cos{\frac{\theta}{q}}\right)
 +m^2 \frac{ \rme^{2\gamma}}{16\pi^2} 
C_+ \left( 1- \cos{\frac{2\theta}{q}} \right)
+\mathcal{O}(m^3 ) ,
\label{eq:Adams-E}
\end{align}
which does not suffer from the UV divergence.

\begin{figure}[tbp]
\centering
  \begin{minipage}[b]{0.47\linewidth}
    \includegraphics[keepaspectratio, width=0.95\textwidth]{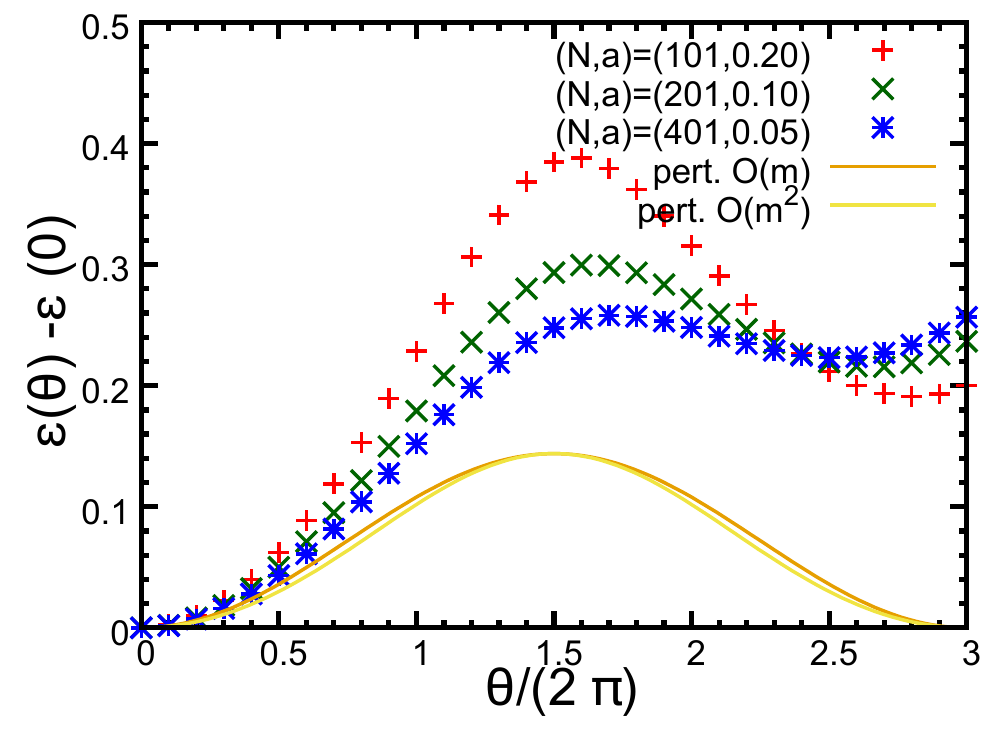}
  \end{minipage}
  \begin{minipage}[b]{0.47\linewidth}
    \includegraphics[keepaspectratio, width=0.9\textwidth]{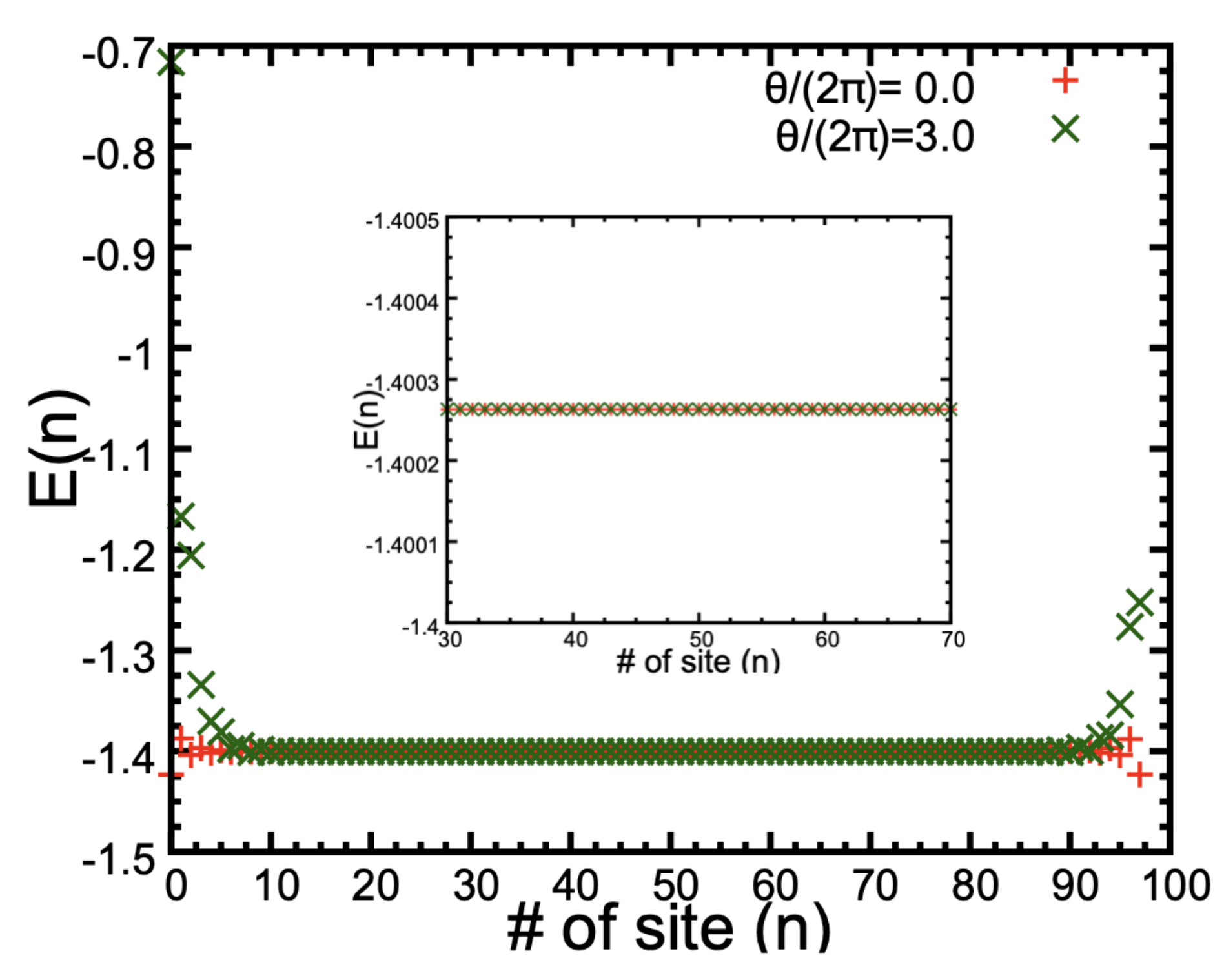}
  \end{minipage}
      \caption{
Results of the averaged energy $\varepsilon(\theta)$ and local energy $E(n)$ 
for the fixed physical volume $L=(N-1)a=20$ and lattice mass $m_{\rm lat}=0.15$. 
(Left) The averaged energy against $\theta /(2\pi )$ for some values of $(N,a)$.
The solid curves denote the results \eqref{eq:Adams-E} of 
the mass perturbation theory at $O(m)$ (orange) and $O(m^2)$ (yellow) in the infinite volume limit.
(Right)~Site dependence of the local energy $E(n)$ 
for $\theta /(2\pi)= 0$ and $3\pi$ with ($N,a$)=($101,0.20$). 
}
      \label{fig:global-E}
\end{figure}
The left panel in Fig.~\ref{fig:global-E} shows 
the averaged energy $\varepsilon(\theta )-\varepsilon(0)$
as a function of $\theta /(2\pi )$. 
We fix the physical volume $L=a(N-1)$ and lattice mass $m_{\rm lat}$ 
as $L=20$ and $m_{\rm lat}=0.15$ 
to demonstrate the typical behavior of the massive charge-$q$ Schwinger model on finite volume with the open boundary condition.
In the figure, we vary values of $(N, a)$ keeping $L$
to grasp a behavior in the continuum limit. 
We see that the simulation data do not show 
the $2\pi q$-periodicity of $\theta$ for all the values of $(N,a)$ and
do deviate from the results \eqref{eq:Adams-E} of 
the mass perturbation theory in the infinite volume limit for non-small $\theta$.
In particular, the data at $\theta/(2\pi) =3$ do not approach zero
as decreasing the lattice spacing $a$.
This is expected behavior in the continuum limit as  continuum theory on a space with boundaries does not have the $2\pi q$-periodicity:
the total ground state energy has a constant term that violates the periodicity due to the emergence of boundary charges
and then the averaged energy slowly converges to the infinite volume limit as $O(1/L)$.
To extract the correct $\theta$ dependence using the averaged energy approximately,
one has to take an infinite volume limit $L\to \infty$ first and then take the continuum limit $a\to 0$, and thus the double extrapolation procedure is necessary (see, for example, Appendix B of Ref.~\cite{Funcke:2019zna}). 

In contrast, the boundary effects should be exponentially small
for the local energy density $E(n)$ away from the boundaries as this is a gapped system.
We can estimate that the effects of the boundaries exponentially decay as a function of $n/\xi$, 
where the correlation length $\xi$ 
is roughly given by 
\begin{\eq}
\xi \approx 
\begin{cases}
\frac{1}{a\mu} =\frac{\sqrt{\pi}}{aqg} & {\rm for\ small}\ m\, (\ll qg),  \cr
\frac{1}{am}  & {\rm for\ large}\ m\, (\gg qg).
\end{cases}
\end{\eq}
In the right panel of Fig.~\ref{fig:global-E}, 
we depict the bare local energy $E_{\mathrm{bare}}(n)$ as a function of site $n$ 
for $\theta=0$ and $\theta/(2\pi)=q$.
We easily see that as going from the boundaries to the center of the lattice, 
both data quickly become almost the same constant
although the values around the boundaries are different.
This implies that 
away from the boundaries, we have the $2\pi q$-periodicity approximately
while more precise checks should be demanded.

\begin{figure}[tbp]
\centering
  \begin{minipage}[b]{0.47\linewidth}
    \includegraphics[keepaspectratio, scale=0.65]{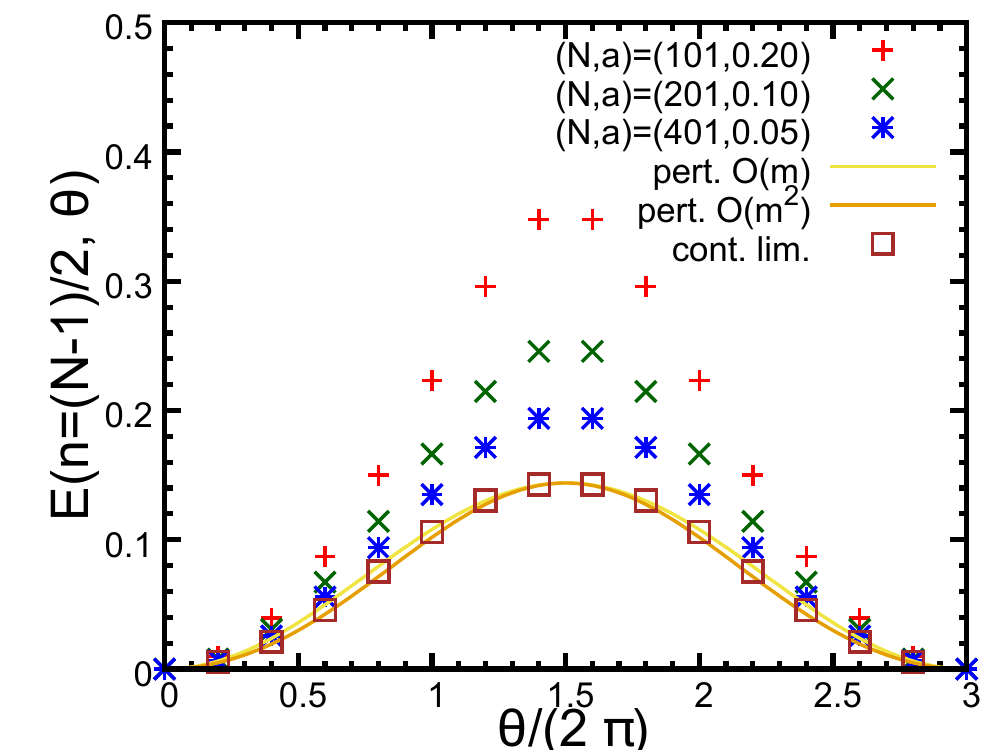}
  \end{minipage}
  \begin{minipage}[b]{0.47\linewidth}
    \includegraphics[keepaspectratio, scale=0.65]{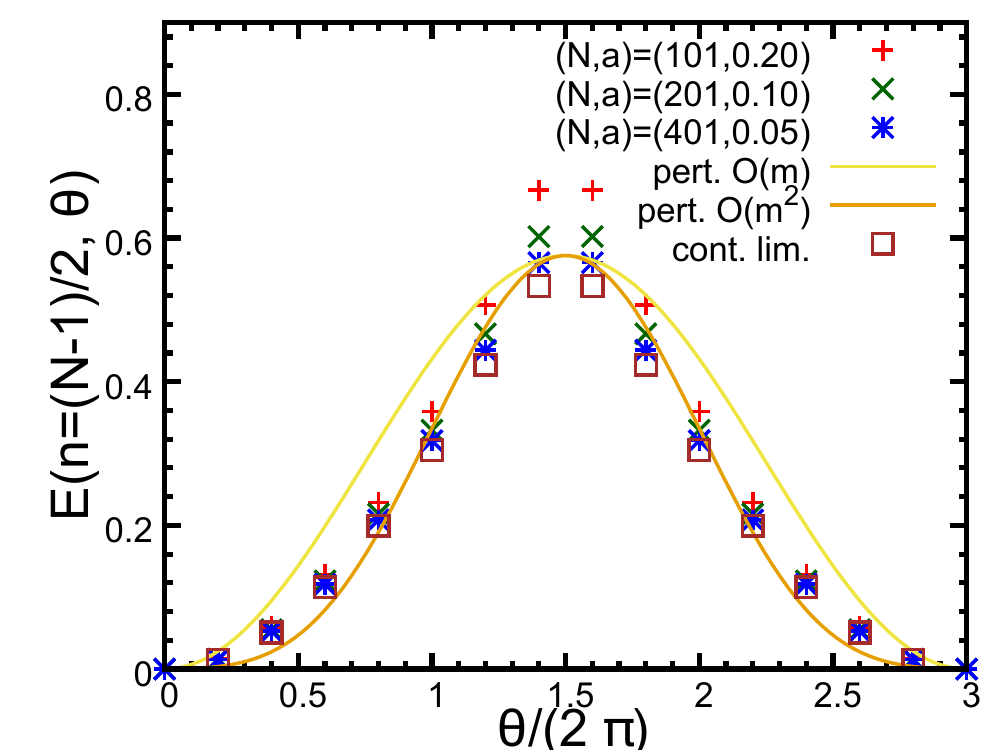}
  \end{minipage}
  \caption{The local energy density $E(n,\theta)$ at the center $n=(N-1)/2$ as a function of $\theta$. 
  Square symbol in each panel denotes the results taken the continuum extrapolation. 
  The mass parameter from left to right panel is $m_{\rm lat}=0.15$ and $0.60$, respectively, and we compare the results with the $\mathcal{O}(m)$ and $\mathcal{O}(m^2)$ mass perturbations.}\label{fig:local-E-theta}
\end{figure}

To see a restoration of the $2\pi q$-periodicity in detail,
let us focus on the local energy $E(n)$ at the central point $n=(N-1)/2$
that should receive the least boundary effects $O(\rme^{-L/2\xi})$.
Now, we shall plot the local energy density at the center as a function of $\theta$ in Fig.~\ref{fig:local-E-theta}.
Here, the lattice fermion mass is $\mlat =0.15$ and $0.60$ from the left to right panels, respectively.
In each panel, the square symbol denotes the result in the continuum limit, where we perform the linear extrapolation of $a$ using three data points; $(N,a)=(101,0.20), (201,0.10)$ and $(401,0.05)$. 
The results clearly show the emergence of $2\pi q$ periodicity for local observables. 
Moreover, the result for $m_{\mathrm{lat}}=0.15$ is consistent with that of mass perturbation, while the higher-order corrections become important for $m_{\mathrm{lat}}=0.60$. 

\begin{figure}[tbp]
\centering
  \begin{minipage}[b]{0.47\linewidth}
    \includegraphics[keepaspectratio, scale=0.65]{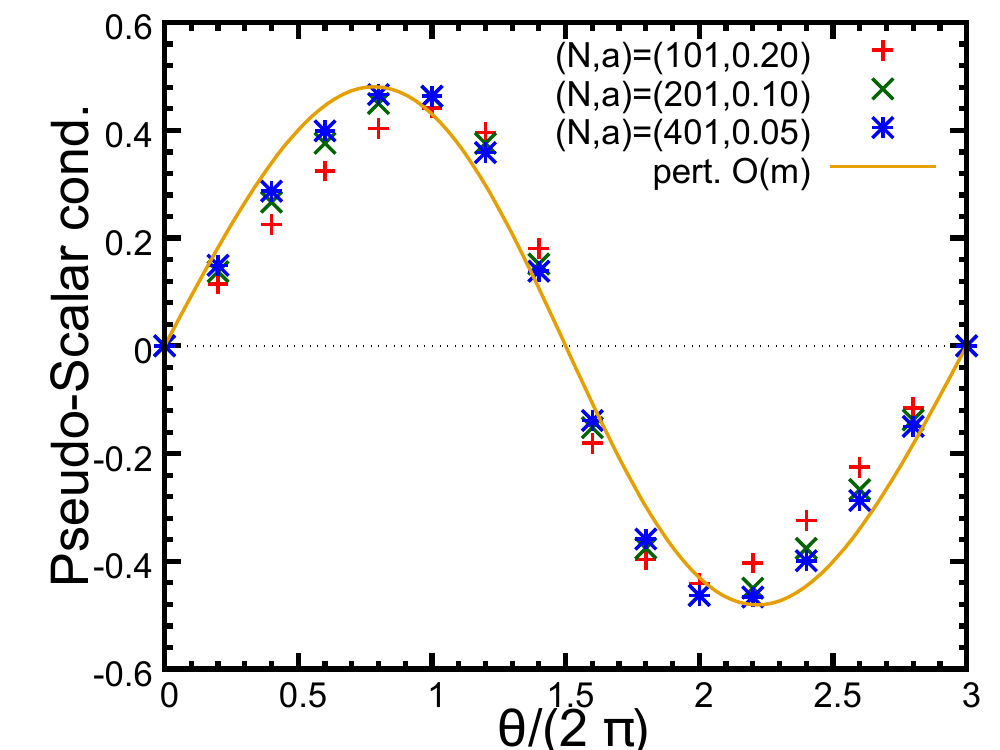}
  \end{minipage}
  \begin{minipage}[b]{0.47\linewidth}
    \includegraphics[keepaspectratio, scale=0.65]{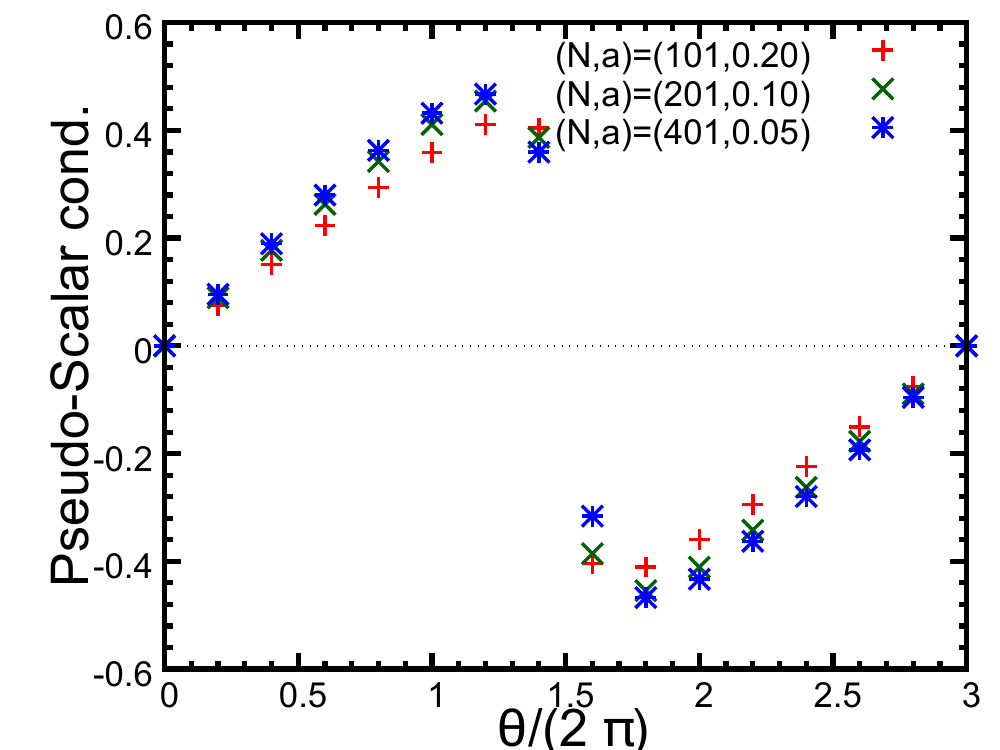}
  \end{minipage}
    \caption{The local pseudo-scalar condensate $P(n)$ at $n=(N-1)/2$ as a function of $\theta$.  
    The mass parameter from left to right panels is $m_{\rm lat}=0.15$ and $0.60$, respectively.
    }
    \label{fig:local-PS-theta}
\end{figure}

Let us also discuss the topological charge density, $\frac{\diff E(\theta)}{\diff \theta}$. Using the ABJ anomaly, it can be related to the expectation value of the pseudo-scalar condensate, 
\begin{equation}
    \frac{\diff \langle E(n,\theta)\rangle}{\diff \theta}=\frac{m}{q}\langle P(n)\rangle. 
    \label{eq:topological_charge}
\end{equation}
The results for the pseudo-scalar condensate is shown in Fig.~\ref{fig:local-PS-theta}, and we can again observe the emergence of the $2\pi q$-periodicity of $\theta$. 
We note that the pseudo-scalar condensate $P(n)$ does not suffer from the UV divergence in our lattice regularization, and thus the results can be compared directly with the analytical results, such as the mass perturbation. 
For $m=0.15$, the results are consistent with one another. 
Detailed study about the continuum limit of the chiral condensates is done in the next subsection. 

\subsection{Chiral condensates and 't Hooft anomaly in the chiral limit}

In the chiral limit, $m=0$, the chiral condensate can be computed analytically\footnote{The mass perturbation of chiral condensates is computed by Ref.~\cite{Adam:1997wt}. 
We note that the bare scalar condensate has UV divergence and it has to be renormalized. 
However, except the chiral limit, the scalar condensate requires the additive renormalization, so we have to match the renormalization scheme to compare the analytical results and numerical computations. 
As this is a nontrivial problem for nonperturbative renormalization, we restrict our attention to the chiral limit to circumvent this issue when we compare the numerical and analytical results. } 
and we find a vacuum expectation value of the chiral condensate operator $O_\pm $ in \eqref{eq:def-chiral} as
\begin{\eq}
S (\theta ) \pm \im\:\! P  (\theta ) 
= \frac{\rme^\gamma qg}{2\pi^{3/2}} e^{\pm\im \frac{\theta}{q} }.
\end{\eq}
As varying $\theta$ in the range $[0,2\pi q]$,
the chiral condensate $(S , P)$ draws a circle around the origin.  

\begin{figure}[tbp]
\centering
  \begin{minipage}[b]{0.49\linewidth}
    \includegraphics[keepaspectratio, scale=0.8]{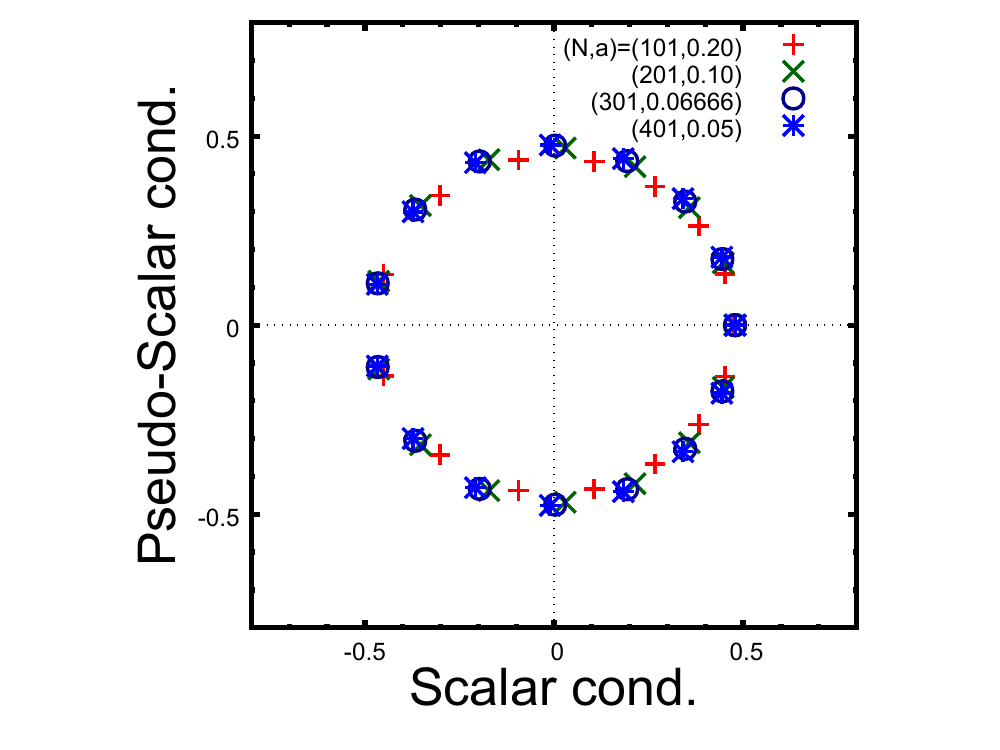}
  \end{minipage}
  \begin{minipage}[b]{0.49\linewidth}
    \includegraphics[keepaspectratio, scale=0.8]{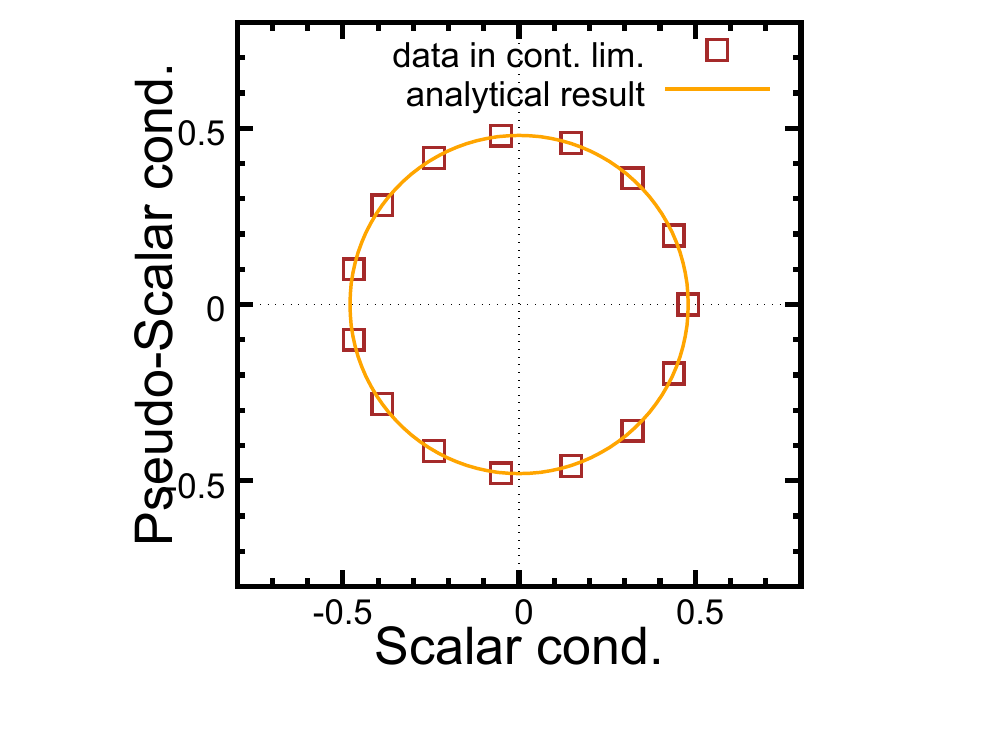}
  \end{minipage}
  \vspace{-1.5em}
      \caption{(Left): $\theta$ dependence of the scalar and pseudo-scalar condensate at $n=(N-1)/2$ site for $m_{\mathrm{lat}}=0.00$ in a fixed volume $L=(N-1)a=20$. (Right): The data in continuum limit (square symbol) and the result of mass perturbation theory. }
     \label{fig:PS-S-M0.00-comp-mass-pert-without-Wloop}
\end{figure}
\begin{figure}[tbp]
\centering
    \includegraphics[keepaspectratio, scale=0.55]{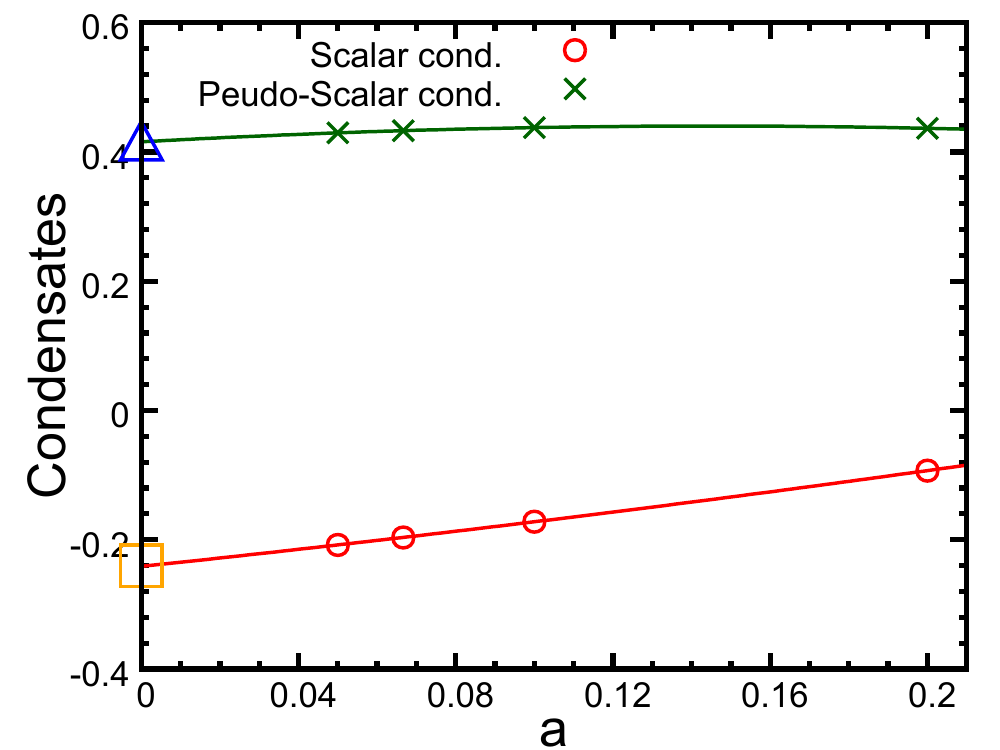}
    \vspace{-0.5em}
    \caption{The continuum extrapolation of the scalar ($\circ$, red) and pseudo-scalar ($\times$, green) for $\theta=2\pi$ using the quadratic function of the lattice constant $a$. The square and triangle symbols at $a=0$ denotes the analytic formula $S=\frac{\mathrm{e}^\gamma qg}{2\pi^{3/2}} \cos{\frac{2\pi}{q}}\simeq -0.240$ and $P=\frac{\mathrm{e}^\gamma qg}{2\pi^{3/2}} \sin{\frac{2\pi}{q}}\simeq 0.415$, respectively.}\label{fig:PS-S-massless}
\end{figure}

Let us confirm the above property by the simulation.
Figure~\ref{fig:PS-S-M0.00-comp-mass-pert-without-Wloop} shows the chiral condensate $(S(\theta),P(\theta))$ in the case of $m_{\mathrm{lat}}=0$ at various values of $\theta$.
We measure the condensates with the interval $\Delta (\theta/2\pi) = 0.2$ in a fixed volume $L=(N-1)a=20$ with $N=101, 201, 301$ and $401$ to take the continuum limit. 
At $\theta=0$, the numerical data are located on the positive real axis, and as $\theta$ increases, the data rotate in the counterclockwise direction.
In the right panel of Fig.~\ref{fig:PS-S-M0.00-comp-mass-pert-without-Wloop}, we show the data in the continuum limit. 
To obtain these data, we perform the quadratic extrapolations of $a$ for $S$ and $P$.
We show its detail in Fig.~\ref{fig:PS-S-massless} taking $\theta=2\pi$ as an example.
As these results indicate, the data in the continuum limit are consistent with the analytical prediction.

\begin{figure}[tbp]
\centering
  \begin{minipage}[b]{0.49\linewidth}
    \includegraphics[keepaspectratio, scale=0.65]{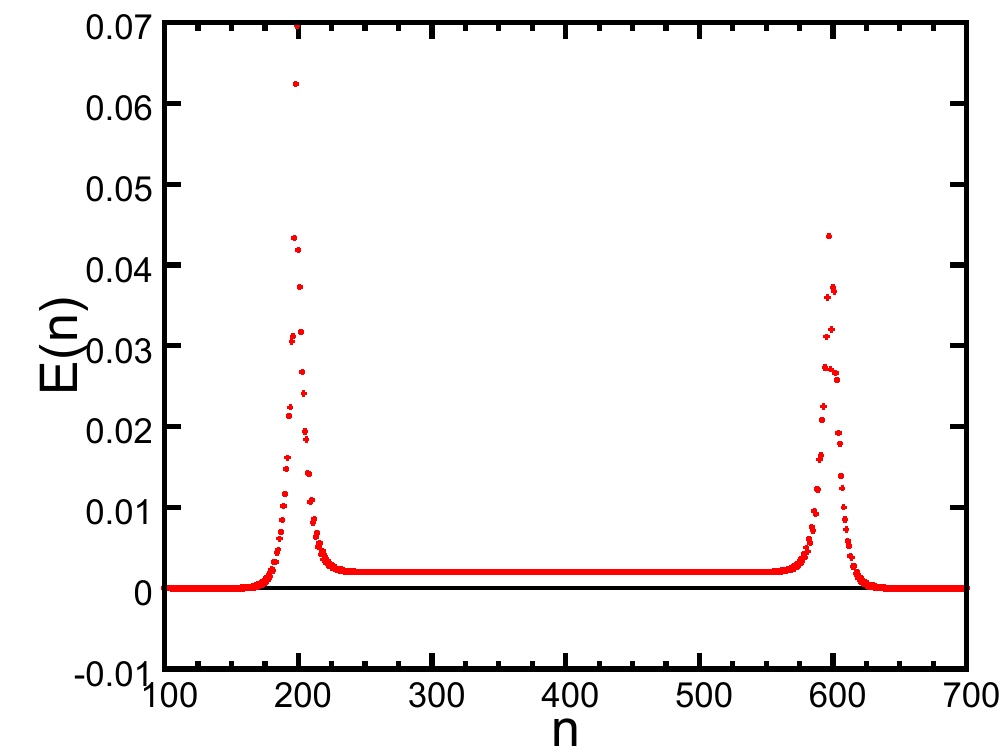}
  \end{minipage}
  \begin{minipage}[b]{0.49\linewidth}
    \includegraphics[keepaspectratio, scale=0.65]{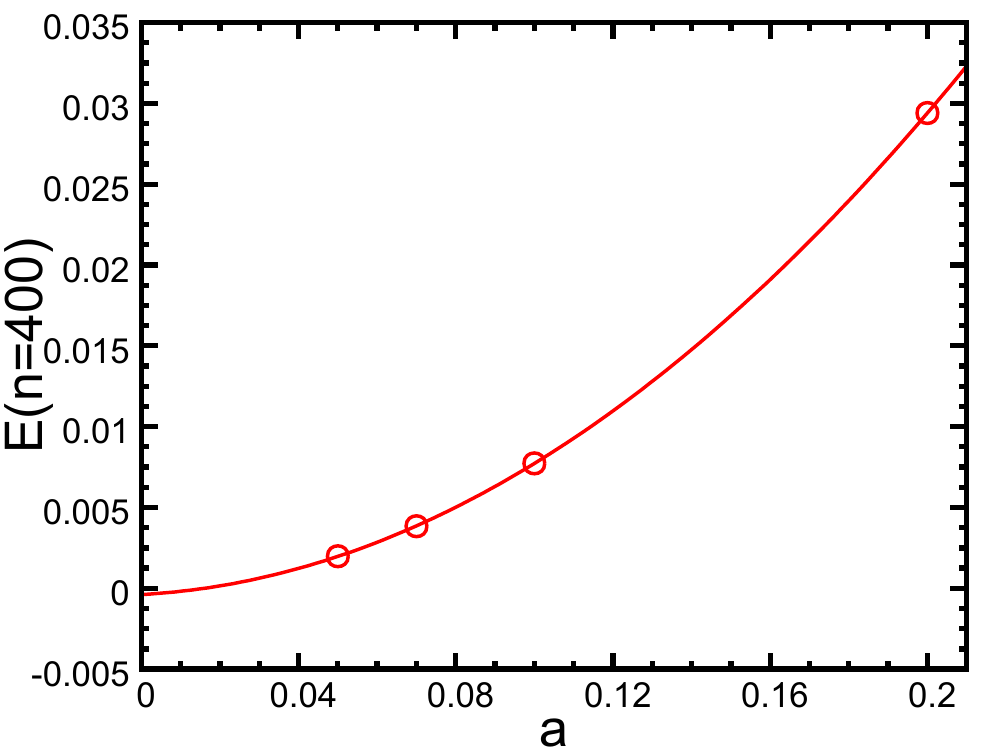}
  \end{minipage}
  \vspace{-1.0em}
      \caption{(Left): Site dependence of the local energy, $E(n)$ with $a=0.05$, $N=801$, $m_{\mathrm{lat}}=0.00$, $\theta =0$
      in the presence of the probe charges with $q_p=2.0$ at $n=200$ and $n=600$.
      (Right): The data in the continuum limit of the local energy at $n=400$. }
     \label{fig:local-E-M0.00-without-Wloop}
\end{figure}

Now let us confirm the effects of the 't Hooft anomaly discussed in Sec.~\ref{sec:Schwinger_review}: 
The analytical computation shows that Wilson loops can be regarded as a generator of the discrete chiral transformation for infrared observers. 
We will investigate whether the phase of the chiral condensation $S(x)+\im\:\! P(x)$ rotates by $2\pi /q$ across the Wilson loop.
For this purpose, we introduce the probe charges $+q_p$ and $-q_p$ with a distance $\hat{\ell} = (N-1)/2$ on the middle of the lattice, 
and then the sites between these probe charges are identified as the inside of the Wilson loop.

Let us first check if the Wilson loop obeys the perimeter law in the chiral limit
meaning that the probe charges are not confined. 
This can be checked by looking at the difference of the local energies between inside and outside of the charges,
which gives the string tension for the confined case and
is expected to be exponentially small as a function of the probe distance, 
namely almost zero, for the screening case. 
In the left panel of Fig.~\ref{fig:local-E-M0.00-without-Wloop}, we put the probe charges at the sites $n=200$ and $n=600$ and plot the local energy density $E(n)$. 
We can see that the local energy density has the sharp peaks around the location of probe charges and it quickly converges to constant values\footnote{When we look the data carefully, our local energy density~\eqref{eq:def-local-energy} has a slightly jagged pattern in addition to the smooth exponential decay in the vicinity of the probe charge. This can be thought of a remnant effect of the staggered fermion, and one may try to remove it by adopting more smooth averaging for the local operator as commented in the footnote for Eq.~\eqref{eq:def-local-energy}. }. 
In the right panel of Fig.~\ref{fig:local-E-M0.00-without-Wloop}, we take the quadratic extrapolation of the local energy at $n=400$ in terms of $a$ to take the continuum limit.
We can see that the local energy inside the Wilson loop is consistent with the one outside the Wilson loop after taking continuum limit and thus the Wilson loop obeys the perimeter law. 
Therefore, for infrared observers, the Wilson loops can be regarded as topological line operators. 

\begin{figure}[tbp]
\centering
    \includegraphics[keepaspectratio, scale=1]{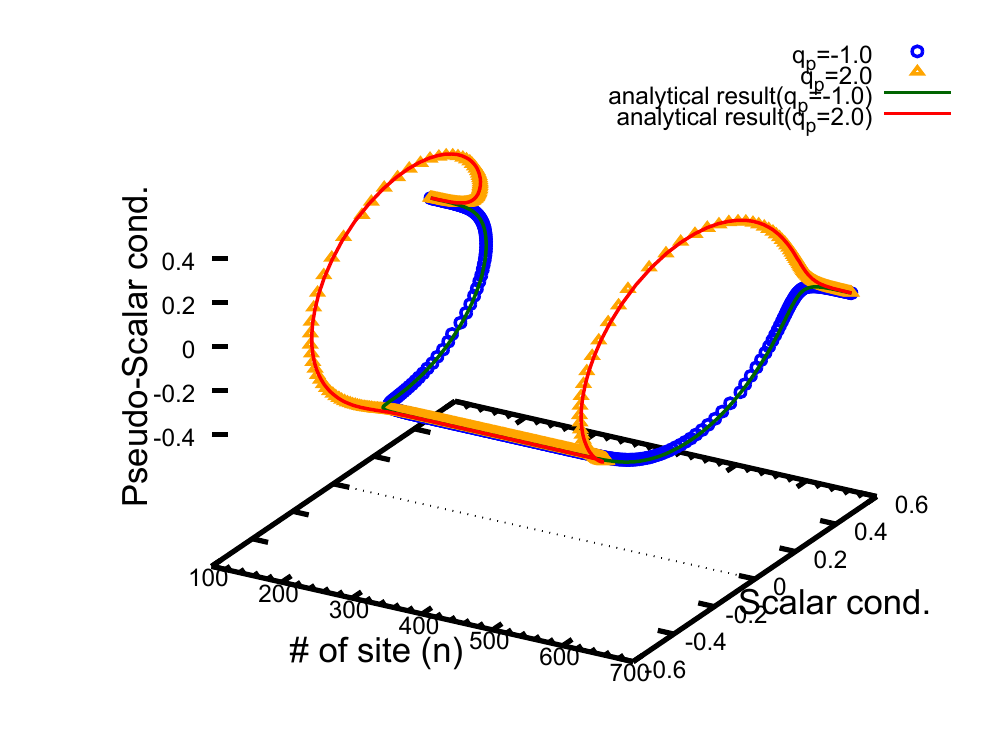}
   \caption{
Position dependence of the local scalar and pseudo-scalar  condensates
for $N=801, a=0.05, m=0, \theta=0$ in the presence of the probe charges at $n=200$ and $600$.
We drop the data at $n< 100$ and $700\le n $ as we are not interested in boundaries.
The circle (blue) and triangle (yellow) symbols denote the probe chargers $q_p=-1.0$ and $q_p=2.0$, respectively. }\label{fig:S-PS-ver-site}
\end{figure}

\begin{figure}[tbp]
\centering
  \begin{minipage}[b]{0.31\linewidth}
    \includegraphics[keepaspectratio, scale=0.57]{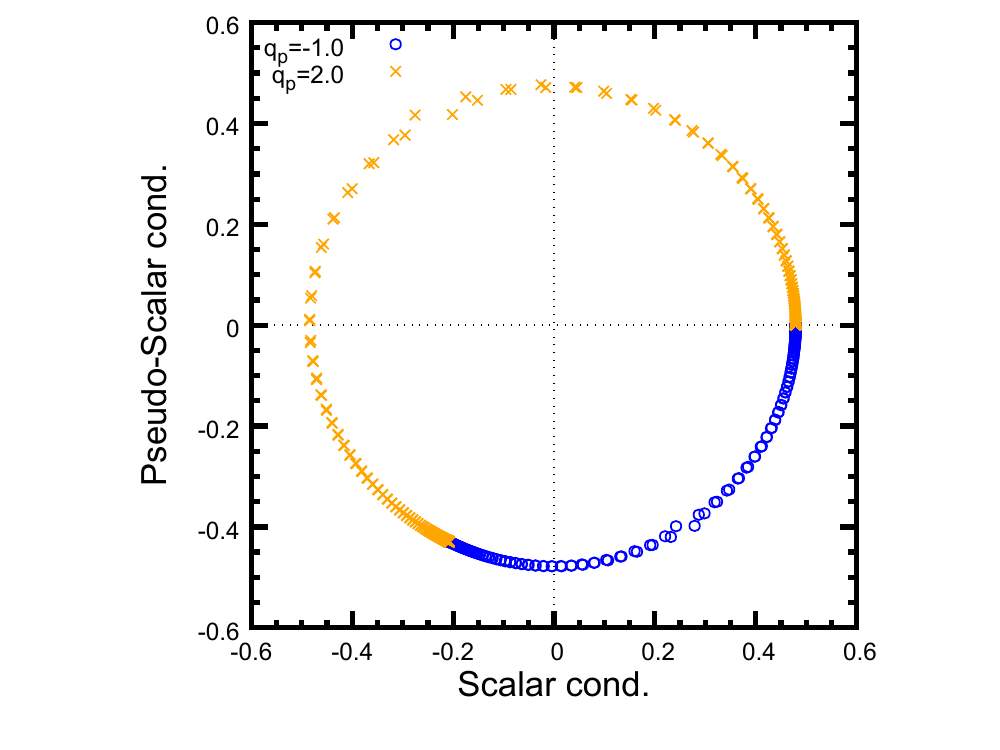}
  \end{minipage}
  \begin{minipage}[b]{0.31\linewidth}
    \includegraphics[keepaspectratio, scale=0.57]{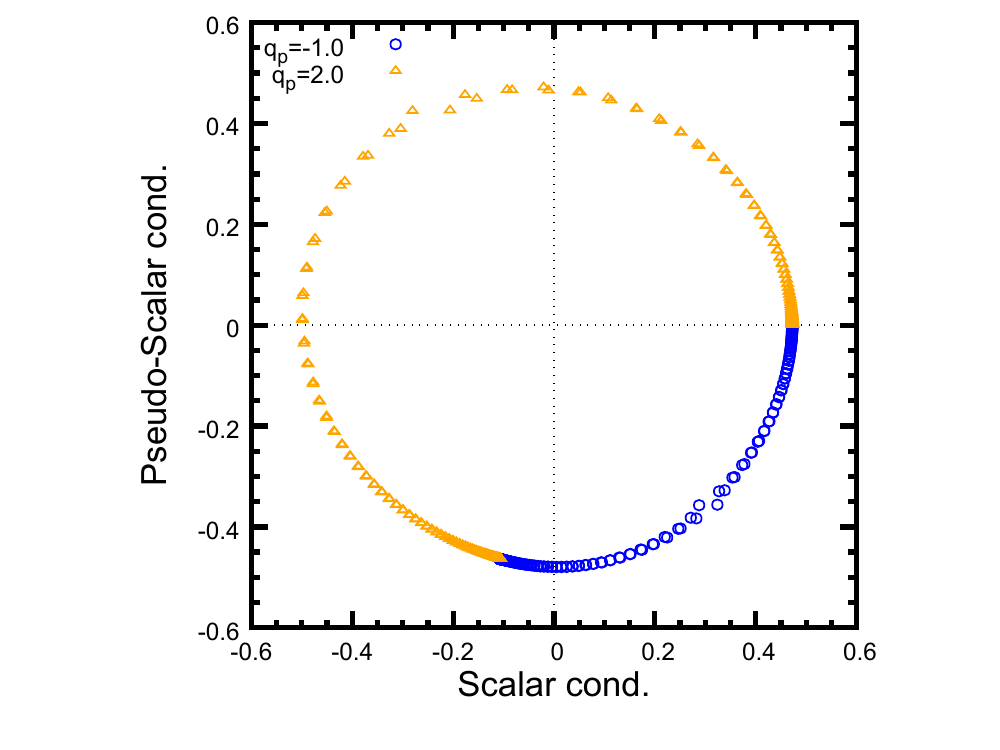}
   \end{minipage}
     \begin{minipage}[b]{0.31\linewidth}
     \includegraphics[keepaspectratio, scale=0.57]{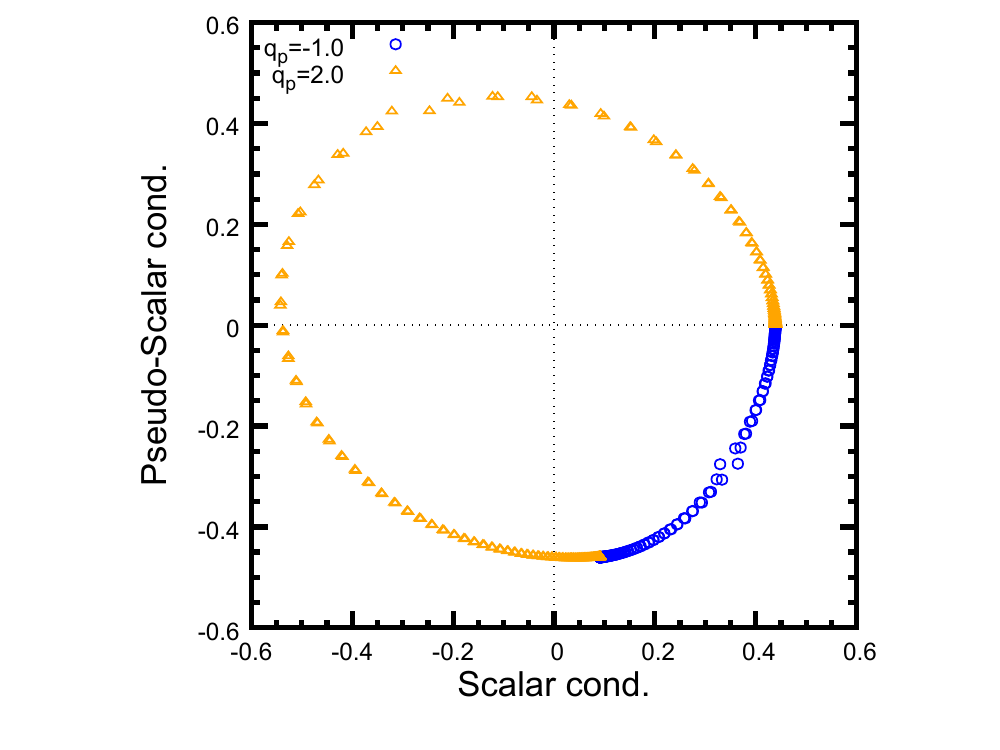}
   \end{minipage}
    \vspace{-1.5em}
     \caption{Scalar and pseudo-scalar condensates in complex plane. The lattice size is $(N,a)=(801,0.05)$ in all plots. The mass parameters are $m_{\mathrm{lat}}=0.00,0.15$ and $0.45$ from left to right panels, respectively.}
    \label{fig:S-PS-mass-deps}
\end{figure}

Figure~\ref{fig:S-PS-ver-site} shows the numerical results of $S(n)$ and $P(n)$ as functions of site $n$.
We also draw the analytical results~\eqref{eq:condensate_with_wilsonloop} based on the bosonization in Fig.~\ref{fig:condensates_3d}.
The numerical results around the probe charges are completely consistent with the result of the analytical computation.
Furthermore, both the condensates with $q_p=-1.0$ and $q_p=2.0$ become almost the same constants away from the Wilson loop, which reflects the fact that the charge-$q$ Schwinger model has the $\mathbb{Z}_q$ 1-form symmetry.
We note that the plateau values inside the Wilson loop ($300 \lesssim n \lesssim 500$) between the numerical data and the analytical results show a small discrepancy, and it comes from the finite $a$ correction that is consistent with Figs.~\ref{fig:PS-S-M0.00-comp-mass-pert-without-Wloop} and \ref{fig:PS-S-massless}.
%
To see the phase of $S(x) +\im P (x)$ clearly,
we take a projection of Fig.~\ref{fig:S-PS-ver-site} to the complex plane and it is shown in the left panel of Fig.~\ref{fig:S-PS-mass-deps}.
The data of $q_p=-1.0$ rotates clockwise as $n$ increases in $n \lesssim 400$ and back to the starting point in $n > 400 $, 
while the one of $q_p=2.0$ rotate counterclockwise in $n \lesssim 400$ and then back to the starting point.
The maximum rotation angle is achieved at $n=400$.
We conclude that the rotation angle by inserting the Wilson loop is $2\pi q_p/q$ as expected from \eqref{eq:1form_symmetry}.

So far we have considered the massless case that is exactly solvable
while it is not easily accessed by the conventional Monte Carlo approach. 
Lastly, let us study how situations are changed when we turn on the nonzero fermion mass $m_{\mathrm{lat}}$, where analytical results are not available. 
We show the results of the site-dependent chiral condensates under the presence of external charges $q_p$ in Fig.~\ref{fig:S-PS-mass-deps} for $m_{\mathrm{lat}}=0.15$ and $0.45$. 
We can see that the rotation angle of chiral condensate becomes shallower as we increase the fermion mass $m_{\mathrm{lat}}$. 
This behavior can be understood from formula~\eqref{eq:topological_charge} for the topological charge. 
Since the topological charge stays finite in the massive fermion limit $m\to \infty$, it suggests that the pseudo-scalar condensate $P$ behaves as $\mathcal{O}(1/m)$ as $m\to \infty$. 
This tendency is already observed also for small fermion mass.

\section{Summary and discussion}
\label{sec:summary}

In this paper, we numerically studied the charge-$q$ Schwinger model by DMRG with a special emphasis on the mixed 't~Hooft anomaly between the $1$-form symmetry and discrete chiral symmetry in the chiral limit.
Upon applying DMRG, we mapped the Schwinger model to the spin chain with the non-local interaction via the Jordan-Wigner transformation, 
and we took the open boundary condition instead of the periodic one to make the Hilbert space finite-dimensional. 
When computing the local energy density or chiral condensate, 
we found that using the local operators significantly reduces the boundary effect compared with the computation of corresponding extensive quantities divided by the volume. 
In particular the local observables exhibited the right periodicity of $\theta$
with a good accuracy in our simulation.
This point should be useful also in simulating more realistic theories such as QCD in future when we take open boundary conditions.
We confirmed that the Wilson loops generate discrete chiral transformations by carefully analyzing the continuum limit. 
When renormalizing the chiral condensate, we find it helpful to relate the lattice fermion mass and the continuum one with the $\mathcal{O}(a)$ correction, $m_{\mathrm{lat}}=m-\frac{q^2 g^2 a}{8}$, suggested by Ref.~\cite{Dempsey:2022nys}.  
Lastly, we also studied how the above property of the chiral condensate operator is changed when we turn on the fermion mass, which is outside of the analytically calculable regime.

For the charge-$q$ Schwinger model, the deconfinement of Wilson loops is the consequence of the discrete chiral symmetry. 
A similar phenomenon is known to occur also for $(1+1)$d adjoint QCD in the chiral limit~\cite{Gross:1995bp, Dalley:1992yy, Bhanot:1993xp}, but it cannot be understood solely from the ordinary chiral symmetry~\cite{Cherman:2019hbq} and requires the presence of noninvertible topological lines~\cite{Komargodski:2020mxz} (For noninvertible symmetries in $(1+1)$d QFTs, see Refs.~\cite{Bhardwaj:2017xup, Chang:2018iay, Thorngren:2019iar}). 
It would be an interesting future study to develop the lattice Hamiltonian formulation for those models and to study them using DMRG.

\subsection*{Acknowledgment}
The authors thank Yuta Kikuchi for early collaboration of this project. 
We also thank Tsuyoshi Okubo for useful comments.
The authors appreciate opportunities for useful discussions during YITP workshops ``Lattice and continuum field theories 2022'' (YITP-W-22-02) and ``A novel numerical approach to quantum field theories'' (YITP-W-22-13). 
M.~H. is supported by MEXT Q-LEAP and JST PRESTO Grant Number JPMJPR2117, Japan.
The work of E.~I. is supported by JSPS KAKENHI with Grant Numbers 
19K03875 and JP18H05407, JST PRESTO Grant Number JPMJPR2113, and the HPCI-JHPCN System Research Project (Project ID: jh220021).
M.~H. and E.~I. are supported by JSPS Grant-in-Aid for Transformative Research Areas (A) JP21H05190.
The work of Y.~T. is partially supported by JSPS
KAKENHI Grant-in-Aid, 22H01218 and
20K22350.
The works of M.~H. and Y.~T. are supported by Center for Gravitational Physics and Quantum Information (CGPQI) at Yukawa Institute.

\appendix
\section{Bond dimension dependence}
\label{sec:app-D-deps}

\begin{figure}[tbp]
\centering
    \includegraphics[keepaspectratio, scale=0.65]{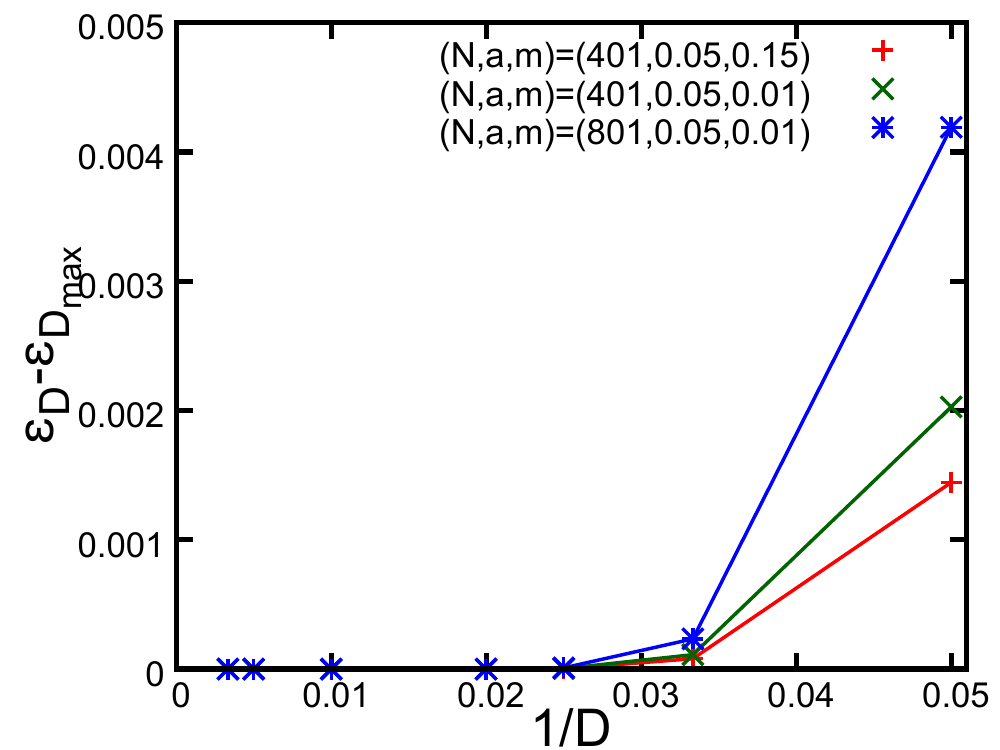}
   \caption{The bond dimension dependence of the ground state energy. The parameter set of simulations is ($N,a,m_{\rm lat.}$)$=$($401,0.05, 0.15$) (red, +), ($401,0.05,0.01$) (green, $\times$) and ($801,0.05,0.01$) (blue, $*$).}\label{fig:D-deps}
\end{figure}

In this appendix, let us investigate systematic errors that come from the fixed bond dimension $D$ in the MPS ansatz~\eqref{eq:MPS-ansatz}. 
Figure~\ref{fig:D-deps} shows the $D$ dependence of the ground state energy, $\varepsilon_D - \varepsilon_{D_{\mathrm{max}}}$ for ($N,a,m_{\rm lat.}$)$=$($401,0.05,0.15$), ($401,0.05,0.01$), and ($801,0.05,0.01$). 
Here, we take the maximal bond dimension as $D_{\mathrm{max}}=300$, and this is used as the reference.
We can see from Fig.~\ref{fig:D-deps} that the result converges for $D\gtrsim 50$, and indeed the difference becomes zero within $10$ digits of precision for $D\ge 50$.

\begin{figure}[tbp]
\centering
   \includegraphics[keepaspectratio, scale=0.65]{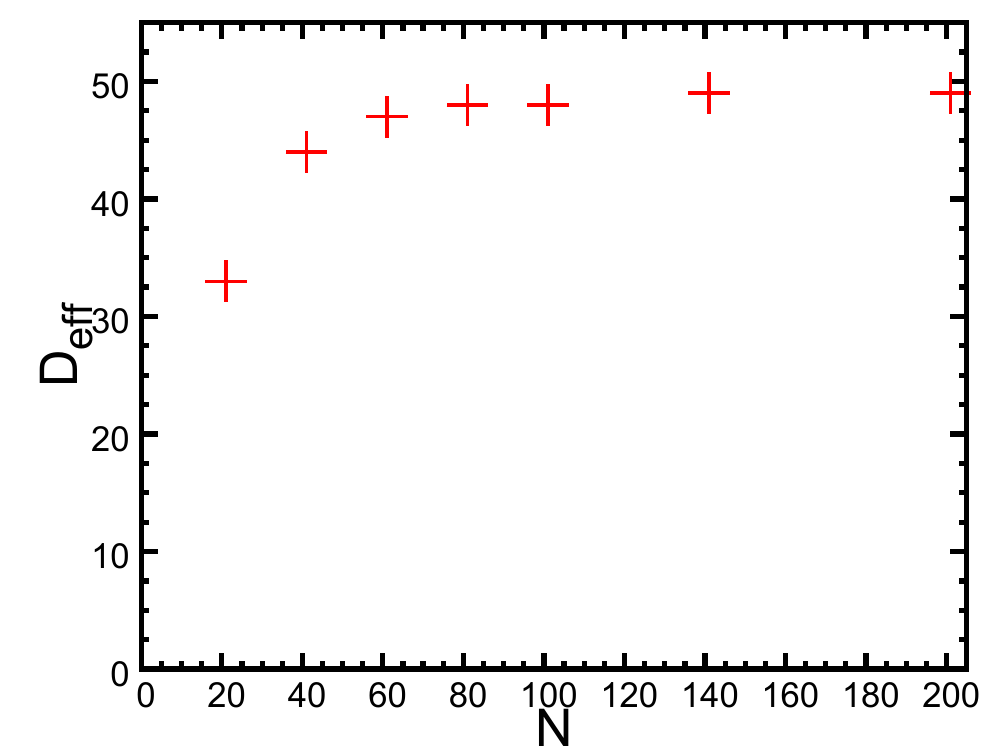}
   \caption{
The system size ($N$) dependence of the effective bond dimension with $a=0.05, m_{\rm lat.}=0.00, \theta/(2\pi)=0.00$ for the charge $q=3$ case at the 10th sweep.}\label{fig:Dmax-massless}
\end{figure}

In order to confirm the system size dependence, we introduce the effective bond dimension $D_{\mathrm{eff}}$: 
In the ITensor, we introduce a cutoff $\epsilon$ in the singular value decomposition and truncate the smallest singular values in a way that the truncation error is less than $\epsilon$. 
We here define $D_{\mathrm{eff}}$ as the number of remaining singular values, and we set $\epsilon=10^{-10}$. 
Although $D_{\mathrm{eff}}$ is not a physical quantity, it gives an upper bound for the entanglement entropy, $S_{\mathrm{EE}}\le \ln D_{\mathrm{eff}}$, so it is helpful to get an idea about how the DMRG works. 
In Fig.~\ref{fig:Dmax-massless}, we plot $D_{\mathrm{eff}}$ as we increase the system size $N$ for $(a,m_{\rm lat.},\theta/2\pi)=(0.05,0,0)$, and $D_{\mathrm{eff}}$ seems to converge around $50$ as $N\to \infty$. 
This convergence implies that the entanglement entropy of the massless Schwinger model obeys the area law, and it is consistent with the fact that this model is gapped due to the ABJ anomaly. 
At generic values of $m$ and $\theta$, the Schwinger model is a gapped system and thus the approximation with fixed bond dimension is expected to be good. 

\begin{figure}[tbp]
\centering
    \includegraphics[keepaspectratio, scale=0.65]{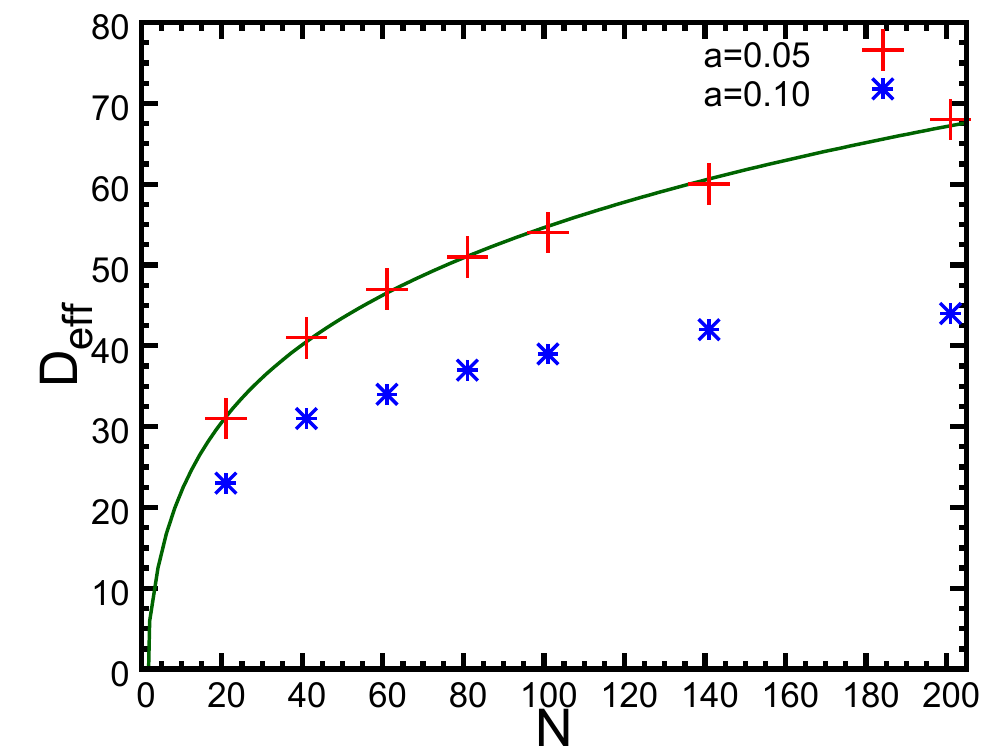}
   \caption{
The system size dependence of the effective bond dimension with $a=0.05$ (red, $+$) and $a=0.10$ (blue, $*$) for the charge $q=3$ case with $m_{\rm lat.}=1.00, \theta/(2\pi)=1.50$ at the 10th sweep.
The green solid line shows a fitting curve by the function $c_1 N^{1/6}+c_2$ with some constants $c_{1,2}$.
}\label{fig:Dmax-critial}
\end{figure}

The charge-$1$ Schwinger model has $2$ vacua for large fermion mass $m>m_*$ at $\theta=\pi$ due to the spontaneous $C$ breaking, and its endpoint $m_*$ is described by the Ising conformal field theory (CFT). 
The critical value of the mass was estimated as $m_*\simeq 0.33$ in Refs.~\cite{Hamer:1982mx, Schiller:1983sj,  Byrnes:2002gj}. 
In the charge-$q$ model, this critical point is mapped to $\theta/(2\pi)=q/2$ and $(m_*/g)\approx 0.33q$, and its local dynamics is again described by the Ising CFT. 
The entanglement entropy of $(1+1)$d CFT has the log correction to the area law and it behaves as $S_{\mathrm{EE}}\sim \frac{c}{3}\ln N$ as $N\to \infty$ with the central charge $c$, and thus $D_{\mathrm{eff}}$ should grow as $D_{\mathrm{eff}}\ge O(N^{c/3})$ at least. 
Although we do not study this case in detail in this paper, let us quickly check the behavior of the effective bond dimension for $a=0.05, m_{\rm lat}=1.00, \theta/(2\pi)=1.50$ with $q=3$. 
The result of the $N$ dependence is shown in Fig.~\ref{fig:Dmax-critial}, and it already shows the tendency for the growth of the bond dimension.
We can see that the behavior of $D_{\mathrm{eff}}$ approximately obeys 
the $N^{1/6}$-law that is consistent with the above expectation with the Ising central charge $c=1/2$. 

In the main text,
the parameter regions are sufficiently far away from this critical point, and the maximal bond dimension is chosen to be sufficiently large values $\in [200,300]$. 
We set the number of sweeps as $20$ and the truncation error cutoff as $10^{-10}$ as the basic DMRG parameters of ITensor. 

\section{The discrete 't~Hooft anomaly on the lattice}
\label{app:discreteanomaly_lattice}

In a recent paper~\cite{Dempsey:2022nys}, it has been found that a part of the discrete 't~Hooft anomaly can be preserved under the lattice regularized Hamiltonian formulation when we take a suitable choice of the lattice mass parameter. 
As in the case of the Lieb-Schultz-Mattis theorem~\cite{Lieb:1961fr, Affleck:1986pq, PhysRevLett.84.1535, Hastings:2003zx}, we can prove rigorously that the ground states should be doubly degenerate when we take the periodic lattice with the even number of sites for the case of even charge $q$, and the one-unit lattice translation plays the essential role there. 
Here, we give a quick review about the lattice realization of the discrete 't~Hooft anomaly for the charge-$q$ Schwinger model. 

\subsection{Discrete \texorpdfstring{$\mathbb{Z}_2$}{Z2} chiral symmetry}

Let us remind the case of the free fermion. 
We consider the periodic lattice with the even number of sites $N\in 2\mathbb{Z}_{>0}$, then the one-flavor naive fermion, 
\begin{align}
    \im w\sum_{n=0}^{N-1}(\chi^\dagger_n \chi_{n+1}-\chi^{\dagger}_{n+1}\chi_n), 
\end{align}
has the dispersion, $\ve(k)=2w\sin\frac{2\pi k}{N}$, and the zero modes locate at the lattice momenta $\frac{2\pi k}{N}=0, \pi$.  
Moreover, they have the opposite chirality and can be regarded as the $(1+1)$d Dirac fermion $\psi(x)$ as we have identified in \eqref{eq:fermion_lattice}. 
This clarifies that the one-unit lattice translation, 
\begin{align}
    \chi_n\mapsto \chi_{n+1},\quad 
    \chi^\dagger_n\mapsto \chi^\dagger_{n+1}, 
\end{align}
acts as the discrete chiral transformation, 
\begin{align}
    \psi(x)\to \rme^{\frac{\pi \im}{2}(-1+\overline{\gamma})}\psi= \overline{\gamma} \psi(x). 
\end{align}
In the continuum formulation, the massless Dirac fermion has the $U(1)_L\times U(1)_R$ chiral symmetry, and the lattice regularization explicitly breaks it down to $U(1)_V\times (\mathbb{Z}_2)_L$. 

The discrete chiral symmetry $(\mathbb{Z}_2)_L$ has a mixed 't~Hooft anomaly with $U(1)_V$ classified by $\mathbb{Z}_2$ and it also has the 't~Hooft anomaly $\Omega_3^{\mathrm{spin}}(B\mathbb{Z}_2)\simeq \mathbb{Z}_8$. 
Although the on-site chiral symmetry is prohibited due to the presence of 't~Hooft anomaly, a part of it can be realized as the local but non-ultralocal symmetry that is related to the lattice symmetry.

To obtain the Schwinger model, we gauge the $U(1)_V$ symmetry by introducing the dynamical gauge field. 
Due to the ABJ anomaly, the continuous axial symmetry is explicitly broken even in the continuum theory. 
As we have seen in Sec.~\ref{sec:Schwinger_review}, however, the charge-$q$ Schwinger model still enjoys the discrete chiral symmetry, given by $(\mathbb{Z}_q)_L$. 
When $q$ is even, 
\begin{align}
    (\mathbb{Z}_2)_L\subset (\mathbb{Z}_q)_L, 
\end{align}
and thus it is a reasonable question to ask if the lattice translation generates the $\mathbb{Z}_2$ subgroup of the discrete chiral symmetry even with the charge-$q$ gauge interaction. 
Ref.~\cite{Dempsey:2022nys} gives the positive answer to this problem.

Although the Hamiltonian~\eqref{eq:lattice_Hamiltonian} with $\mlat=0$ may look to have the one-unit translation invariance, we should note that the symmetry operation should be also consistent with the canonical commutation relations~\eqref{eq:commutation_relation} and with the Gauss law constraint~\eqref{eq:Gauss_law}:
\[
L_n -L_{n-1} 
= q \Biggl[ \chi_n^\dag \chi_n -\frac{1-(-1)^n}{2}  \Biggr].
\]
In the case of the Schwinger model, the naive one-unit lattice translation is inconsistent with the Gauss law due to the presence of the staggering constant, $\frac{1-(-1)^n}{2}$, on the right-hand-side of \eqref{eq:Gauss_law}. 
To keep the Gauss law~\eqref{eq:Gauss_law} intact, we define the one-unit lattice translation of $L_n$ as 
\begin{align}
    L_n\mapsto L_{n+1}+q\frac{1-(-1)^{n+1}}{2}. 
    \label{eq:oneunit_Ln}
\end{align}
Then, the left-hand-side of \eqref{eq:Gauss_law} transforms as $L_n-L_{n-1}\mapsto L_{n+1}-L_n+q(-1)^n$, while the right-hand-side becomes $q[\chi^\dagger_n \chi_n-\frac{1-(-1)^{n}}{2}]\mapsto q[\chi^\dagger_{n+1} \chi_{n+1}-\frac{1-(-1)^{n+1}}{2}]+q(-1)^n$.
We note that the additive constant $q\frac{1-(-1)^{n+1}}{2}$ is an integer, so it does not change the spectrum of $L_n$ and thus this is a well-defined operation. 
We can also readily confirm that it does not change the canonical commutation relation. 

Thus, the remaining task is to establish the invariance of the Hamiltonian. The gauge kientic term transforms as 
\begin{align}
    J\sum_{n}\left(L_n+\frac{\theta_0}{2\pi}\right)^2 
    &\mapsto J\sum_{n}\left(L_{n+1}+q\frac{1-(-1)^{n+1}}{2}+\frac{\theta_0}{2\pi}\right) \nonumber\\
    &= J\sum_{n}\left(L_n+\frac{\theta_0+q\pi}{2\pi}-\frac{(-1)^n q}{2}\right)^2\nonumber\\
    &=J\sum_n\left(L_n+\frac{\theta_0+q\pi}{2\pi}\right)^2\nonumber\\
    &\quad -\frac{qJ}{2}\sum_n (-1)^n\left\{(L_{n}-L_{n-1})+q\frac{1-(-1)^n}{2}\right\}. 
\end{align}
Using the Gauss law~\eqref{eq:Gauss_law}, the last term takes the same form 
as the mass term on the physical Hilbert space. 
Therefore, we have 
\begin{align}
    &\qquad J\sum_{n}\left(L_n+\frac{\theta_0}{2\pi}\right)^2+ \mlat \sum_{n}(-1)^n \chi^\dagger_n \chi_n\nonumber\\
    &\mapsto J\sum_{n}\left(L_n+\frac{\theta_0+q\pi}{2\pi}\right)^2-\left(\mlat +\frac{q^2 J}{2}\right)\sum_{n}(-1)^n \chi^\dagger_n \chi_n.
\end{align}
In particular, when we choose the lattice mass parameter as~\cite{Dempsey:2022nys} 
\begin{align}
    \mlat =-\frac{q^2 J}{4}=-\frac{q^2 g^2 a}{8}, 
\end{align}
the one-unit lattice translation relates the Hamiltonian at $\theta_0$ and $\theta_0+q\pi$: 
\begin{align}
    H_{\theta_0}\mapsto H_{\theta_0+q\pi}=\mathcal{T} H_{\theta_0}\mathcal{T}^{-1}, 
\end{align} 
where $\mathcal{T}$ is the one-unit lattice translation defined above. 
When $q$ is even, we can use the unitary operator, $W^{q/2}(S^1)=(\prod_n U_n)^{q/2}$, to obtain the original Hamiltonian, 
\begin{align}
    H_{\theta_0}
    &=W^{q/2}(S^1) H_{\theta_0+q\pi} (W^{q/2}(S^1))^{-1}\nonumber\\
    &=(W^{q/2}(S^1) \mathcal{T})\, H_{\theta_0}\, (W^{q/2}(S^1)\mathcal{T})^{-1},
\end{align}
and thus the one-unit lattice translation $\mathcal{T}$ becomes the good symmetry operation by associating it with the above unitary transformation. 

\subsection{Exact \texorpdfstring{$2$}{2}-fold degeneracy on the periodic even lattice with even \texorpdfstring{$q$}{q}}

For the even-charge lattice Schwinger model on the periodic even lattice, we have the $(\mathbb{Z}_2)_L$ chiral symmetry with the suitable choice of the mass parameter, $m=-\frac{q^2 g^2 a}{8}$. 
In the continuum theory, there is the mixed anomaly between $\mathbb{Z}_q^{[1]}$ and $(\mathbb{Z}_q)_L$, and it still exists even if we break the chiral symmetry to $(\mathbb{Z}_2)_L\subset (\mathbb{Z}_q)_L$ when $q$ is even. 
As a result of the anomaly matching, the chiral symmetry should be spontaneously broken. 

Let us prove the lattice counterpart of this statement to conclude the double degeneracy of the ground states. 
For this purpose, we note that the $\mathbb{Z}_q^{[1]}$ symmetry is generated by 
\begin{align}
    \rme^{\frac{2\pi\im}{q}L_n}. 
\end{align}
Due to the Gauss law, this operator does not depend on the spatial sites $n$, and we can easily check that it commutes with the Hamiltonian. 
Moreover, $(\rme^{\frac{2\pi\im}{q}L_n})^q=\rme^{2\pi \im L_n}$ gives the large gauge transformation, and thus it should be the identity, $1$, on the physical Hilbert space $\mathcal{H}$. 
As a result, the physical Hilbert space decomposes into the $q$ distinct sectors, 
\begin{align}
    \mathcal{H}=\bigoplus_{k=1}^{q} \mathcal{H}_k,
\end{align}
where 
\begin{align}
    \mathcal{H}_k=\{\psi\in \mathcal{H}\, | \, \rme^{\frac{2\pi\im}{q}L_n}\psi=\rme^{\frac{2\pi \im k}{q}}\psi\}. 
\end{align}
As the discrete chiral symmetry is generated by $W^{q/2}(S^1)\mathcal{T}$, let us compute its commutation relation with the $1$-form symmetry generator:
\begin{align}
    \rme^{\frac{2\pi\im}{q}L_n} (W^{q/2}(S^1)\mathcal{T}) &= - (W^{q/2}(S^1)\mathcal{T})\, \rme^{\frac{2\pi\im}{q}L_{n-1}} \nonumber\\
    &= - (W^{q/2}(S^1)\mathcal{T})\, \rme^{\frac{2\pi\im}{q}L_{n}}.  
\end{align}
Therefore, the Hamiltonian has the same energy spectrum on $\mathcal{H}_k$ and $\mathcal{H}_{k+q/2}$. 
This shows that the whole energy spectrum must be two-fold degenerate, and, in particular, so is the ground state. 
Assuming the presence of the mass gap, we can conclude the spontaneous breaking of $(\mathbb{Z}_2)_L$ chiral symmetry in the thermodynamic limit.

\bibliographystyle{utphys}
\bibliography{quantum_computation,./QFT}

\end{document}